\documentclass[aps,reprint,onecolumn,pra,showpacs,superscriptaddress,floatfix,notitlepage,nofootinbib]{revtex4-1}
\usepackage{amssymb,amsmath,amsfonts} 
\usepackage{epsfig,graphicx}

\newcommand{\ket}[1] {\ensuremath{\left| #1 \right>}}
\newcommand{\bra}[1] {\ensuremath{\left< #1 \right|}}

\newcommand{\braket}[2] {\ensuremath{\left< #1 | #2 \right>}}
\newcommand{\expected}[3] {\ensuremath{\left< #1 \right| #2 \left| #3 \right>}}

\newcommand{\coef}[2] {\ensuremath{\sqrt{\frac{#1}{#2}}}}

\newcommand{\head}{\ensuremath{\ket{\text{head}}_R}}
\newcommand{\tail}{\ensuremath{\ket{\text{tail}}_R}}

\newcommand{\arriba}{\ensuremath{\ket{\uparrow}_S}}
\newcommand{\abajo}{\ensuremath{\ket{\downarrow}_S}}
\newcommand{\dcha}{\ensuremath{\ket{\rightarrow}_S}}

\begin{document}
\title{Decoherence framework for Wigner's friend experiments}

\author{Armando Rela\~{n}o} \affiliation{Departamento de Estructura de
  la Materia, F\'{\i}sica T\'ermica y Electr\'onica, and GISC,
  Universidad Complutense de Madrid, Av. Complutense s/n, 28040
  Madrid, Spain} \email{armando.relano@fis.ucm.es}
\begin{abstract}
  The decoherence interpretation of quantum measurements is applied to
  Wigner's friend experiments. A framework in which all the
  experimental outcomes arise from unitary evolutions is
  proposed. Within it, a measurement is not completed until an
  uncontrolled environment monitorizes the state composed by the
  system, the apparatus and the observer. The (apparent) wave-function
  collapse and the corresponding randomness result from tracing out
  this environment; it is thus the ultimate responsible for the
  emergence of definite outcomes. Two main effects arise from this
  fact. First, external interference measurements, trademark of
  Wigner's friend experiments, modify the memory records of the
  internal observers; this framework provides a univocal protocol to
  calculate all these changes. Second, it can be used to build a
  consistent scenario for the recenly proposed extended versions of
  the Wigner's friend experiment. Regarding [D. Frauchiger and
    R. Renner, {\em Quantum theory cannot consistently describe the
      use of itself}, Nat. Comm. {\bf 9}, 3711 (2018)], this framework
  shows that the agents' claims become consistent if the changes in
  their memories are properly taken into account. Furthermore, the
  particular setup discussed in [C. Brukner, {\em A no-go theorem for
      observer-indepdendent facts}, Entropy {\bf 20}, 350 (2018)]
  cannot be tested against the decoherence framework, because it does
  not give rise to well-defined outcomes according to this
  formalism. A variation of this setup, devised to fill this gap, makes it
  possible to assign joint truth values to the observations made by
  all the agents. This framework also narrows down the requisites for
  such experiments, making them virtually impossible to apply to
  conscious (human) beings. Notwithstanding, it also opens the door to
  future relizations on quantum machines.
\end{abstract}

\maketitle

\section{Introduction}

In 1961, Eugene Wigner proposed a thought experiment to show that a
conscious being must have a different role in quantum mechanics than
that of an inanimate device \cite{Wigner:61}. This experiment consists of two
observers playing different roles. The first one, Wigner's friend,
performs a measurement on a particular quantum system in a closed
laboratory; as a consequence of it, she observes one of the possible
outcomes of her experiment. The second one, Wigner himself, measures
the whole laboratory from outside. If quantum theory properly accounts
for what happens inside the laboratory, Wigner observes that both his
friend and the measured system are in an entangled superposition
state. Hence, the conclusions of both observers are incompatible. For
Wigner's friend, the reality consists in a definite state equal to one
of the possible outcomes of her experiment; for Wigner, it
consists in a superposition of all these possible outcomes.

Since then, a large number of discussions, interpretations and
extensions have been done. Among them, this work focus on a recent extended
version of this experiment, from which two different no-go theorems
have been formulated. The first one shows that different agents,
measuring on and reasoning over the same quantum system, are bound to
get contradictory conclusions \cite{Renner:18}. The second one
establishes that it is impossible to assign join truth values to the
observations made by all the agents \cite{Bruckner:18}. This extended
version of the Wigner's friend experiment consists of two closed
laboratories, each one with an observer inside, and two outside
observers dealing with a different laboratory. All the
measurements are performed on a pair of entangled quantum systems,
each one being measured in a different laboratory. An experiment to
prove the second no-go theorem has been recently done
\cite{Proietti:19}.

The key point of the original and the extended versions of the
Wigner's friend experiment is the quantum treatment of the
measurements performed inside the closed laboratories. It is assumed
that Wigner's friend observes a definite outcome from her experiment,
but the wavefunction of the whole laboratory in
which she lives remains in an entangled superposition state. This is
somehow in contradiction with the spirit of the Copenhaguen
interpretation, since the measurement does not
entail a non-unitary collapse. Its main shortcoming is not providing a
specific procedure to determine whether a proper measurement has been
performed. It is not clear at all whether an agent has observed a
definite outcome, or just a simple quantum correlation, implying no
definite outcomes, has been crafted. But, at the same time, it can be
useful in the era of quatum technologies, because it can describe the
evolution of a quantum machine able to perform experiments, infer
conclusions from the outcomes, and act as a consequence of them.

The aim of this work is to provide a framework which keeps the quantum
character of all the measurements, while supplying a mechanism for the
(apparent) wavefunction collapse that the agents perceive. This is
done by means of the decoherence interpretation of quantum
measurements \cite{Zurek:03}, whose origin comes back to almost forty
years ago \cite{Zurek:81}. The key element of this interpretation is
that a third party, besides the measured system and the measuring
apparatus, is required to complete a quantum measurement. It consists
in an uncontrolled environment, which cannot be the object of present
or future experiments, and which is the ultimate responsible of the
emergence of definite outcomes, and the (apparent) wavefunction
collapse. Hence, the laboratory in which Wigner's friend lives must
include three different objects: the measured system, the measuring
apparatus, and the uncontrolled environment ---a quantum machine
performing such an experiment must contain a set of qbits making up
the measuring apparatus and the computer memory, and a second set of
qbits forming the environment; the Hamiltonian of the complete
machine, including all these qbits, is supossed to be known. The
decoherence formalism establishes that this environment, not present
in standard Wigner's friend setups
\cite{Wigner:61,Renner:18,Bruckner:18,Proietti:19}, determines to
which states the memory of Wigner's friend collapses, and therefore
which outcomes are recorded by her. And, at the same time, it
guarantees the unitary evolution of the whole laboratory, making it
possible for Wigner to observe the system as an entangled
superposition of his friend, the measuring apparatus, {\em and} the
environment. Notwithstanding, our aim is not to support this framework
against other possibilities, like wave-function collapse theories, for
which the collapse is real and due to slight modifications in the
quantum theory that only become important for large systems
\cite{Bassi:13}; or recently proposed modifications of the Born rule
\cite{Baumann:18}.  We just intend to show that: {\em (i)} this
framework provides a univocal protocol to calculate the state of the
memories of all the agents involved in the experiment, at any time;
{\em (ii)} it rules out all the inconsistencies arising from the
standard interpretations of Wigner's friend experiments, and {\em
  (iii)} it narrows down the circumstances under which such
experiments can be properly performed. We can trust in future
experiments involving quantum {\em intelligent} machines to determine
which is the correct alternative ---if any of these.

Our first step is to build a simple model for the interaction between
the measuring apparatus and the environment. This model allows us to
determine the properties of the interaction and the size of the
environment required to give rise to a proper measurement, as
discussed in \cite{Zurek:03}. Therefore, it can be used to build a
quantum machine to perform Wigner's friend experiments. Then, we
profit from it to discuss the original Wigner's friend experiment
\cite{Wigner:61}, and the no-go theorems devised in
\cite{Renner:18,Bruckner:18}. We obtain the following
conclusions. First, the external interference measurement that Wigner
performs on his friend changes her memory record. This change can be
calculated, and its consequences on further measurements can be
exactly predicted. Second, the decoherence framework rules out all the
inconsistencies arising from the usual interpretations of these
experiments. Finally, it also establishes restrictive requirements for
such experiments.

To avoid all the difficulties that conscious (human) beings entail,
all the observers are considered quantum machines, that is, devices
operating in the quantum domain, and programmed with algorithms
allowing them to reach conclusions from their own observations. This
choice facilitates the challenge of the experimental verification (or
refutation) of the results that the decoherence framework provides, against, for example, predictions of wave-function collapse models
\cite{Bassi:13}, or modifications of the Born rule
\cite{Baumann:18}. Within this spirit, all the Hamiltonians discussed
throughout this paper must be understood as fundamental parts
of quantum machines dealing with Wigner's friend
experiments; the Hamiltonians modelling the algorithms used by any
particular setup are far beyond the scope of this paper.

The paper is organised as follows. Sec. \ref{sec:decoherencia} is
devoted to the decoherence interpretation of quantum measurements. A
simple numerical model is proposed to guide all the discussions. In Sec. \ref{sec:wigner}, the original Wigner's friend
experiment is studied in terms of the decoherence framework. A
numerical simulation is used to illustrate its most significative
consequences.  In Sec. \ref{sec:renner} the consistency of the quantum
theory is discussed, following the argument devised in
\cite{Renner:18}. Sec. \ref{sec:bruckner} refers to the possibility of
assigning joint truth values to all the measurements in an extended
Wigner's friend experiments, following the point of view publised in
\cite{Bruckner:18}. Finally, conclusions are gathered in
Sec. \ref{sec:conclusiones}.

\section{Decoherence framework}
\label{sec:decoherencia}

The first aim of this section is to review the decoherence
formalism. We have chosen the examples and adapted the notation to
facilitate its application to Wigner's friend experiments. After this
part is completed, we propose a Hamiltonian model giving rise
to the definite outcomes observed by any agents involved in any
quantum experiment, and we explore its consequences by means of
numerical simulations.

\subsection{Decoherence interpretation of quantum measurements}

In all the versions of Wigner's friend experiments, the protocol
starts with a measurement performed by a certain agent, $I$. Let us
consider that a single photon is the object of such measurement, and
let us suposse that the experiment starts from the following initial
state,
\begin{equation}
  \ket{\Psi} = \coef{1}{2} \left( \ket{h} + \ket{v} \right),
  \label{eq:decoherencia}
\end{equation}
where $\ket{h}$ denotes that it is horizontally polarised, and
$\ket{v}$, vertically polarised.

The usual way to model a quantum measurement consists in a unitary evolution,
given by the Hamiltonian that encodes the dynamics of the system and
the measuring apparatus. It transforms the initial state, in which
system and apparatus are uncorrelated, onto a final state in which the
system and the apparatus are perfectly correlated
\begin{equation}
  \frac{1}{\sqrt{2}} \left( \ket{h} + \ket{v} \right) \otimes \ket{A_0} \longrightarrow \frac{1}{\sqrt{2}} \left(\ket{h} \otimes \ket{A_h} + \ket{v} \otimes \ket{A_v} \right),
  \label{eq:measurement1}
\end{equation}
where $\ket{A_0}$ represents the state of the apparatus before the
measurement, $\braket{A_h}{A_v}=0$, and
$\braket{A_h}{A_h}=\braket{A_v}{A_v}=1$.  In \cite{Proietti:19}, an
ancillary photon plays the role of the apparatus. In general, such a
measurement can be performed by means of a C-NOT gate. As the choice
of $\ket{A_0}$ is arbitrary, we can consider that $\ket{A_0} \equiv
\ket{A_h}$, and thus the corresponding Hamiltonian is given by
\begin{equation}
  \begin{split}
    H &= \frac{g}{2} \ket{v}\bra{v} \otimes \\ &\left[ \ket{A_h}\bra{A_h} +  \ket{A_v}\bra{A_v} -  \ket{A_v}\bra{A_h} -  \ket{A_h}\bra{A_v}  \right],
    \end{split}
  \label{eq:medida}
\end{equation}
where $g$ is a coupling constant. This Hamiltonian performs
Eq. (\ref{eq:measurement1}), if it is applied during an interaction
time given by $g \tau=\pi/2$ \cite{Zurek:03}. The resulting state, which we denote
\begin{equation}
  \ket{\Psi} = \coef{1}{2} \left( \ket{h} \ket{A_h} + \ket{v} \ket{A_v} \right)
  \label{eq:measurement2}
\end{equation}
for simplicity, entails that if the photon has horizontal
polarisation, then the apparatus is in state $\ket{A_h}$, and if the
photon has vertical polarisation, then the apparatus is in state
$\ket{A_v}$. That is, it is enough to observe the apparatus to know
the state of the photon.

As we have just pointed out, this is the usual description of a
quantum measurement. Once the state \eqref{eq:measurement2} is fixed,
the measurement is completed, and the only remaining task is to
interpret the results. This is precisely what is done in the
experimental facility discussed in \cite{Proietti:19}. In both the
original and the extended versions of Wigner's friend experiments,
the interpretation is the following. The observer inside the
laboratory, $I$, sees that the outcome of the experiment is either $h$
or $v$, with probability $1/2$, following the standard Born rule; it
sees the reality as consisting in a definite state corresponding
either to $\ket{h}$ or $\ket{v}$, according to the information it
has gathered. Even more, it can write that its observation has been
completed, making possible for an external obsever, $E$, to know that
$I$ is seeing a definite outcome,
\begin{equation}
  \ket{\Psi'_1} = \left[ \frac{1}{\sqrt{2}} \left( \ket{h} \ket{A_h} + \ket{v} \ket{A_v} \right) \right] \otimes \ket{\text{Observation}}.
  \label{eq:measurement3}
\end{equation}
This implies that $I$ has observed a definite outcome, whereas the
whole laboratory in which it lives remains in a superposition state
that can be observed by $E$, despite knowing that $I$ sees the photon
either in horizontal or vertical polarisation, and not in such a
superposition state.

This conclusion is the basis of all the versions of the Wigner's
friend experiment. Notwithstanding, it suffers from two important
shortcomings. The first one is that the complete laboratory consists
just in the measured system and the measuring apparatus. Hence, there
is no place for a quantum device able to act as a consequence of its
measurement ---the reasonings to infer contradictory conclusions, as
discussed in \cite{Renner:18}, require a complex machine, not just a
qbit signaling whether the measured photon is vertically or
horizontally polarised. Therefore, as is pointed out in
\cite{Proietti:19}, the consideration of Eq. (\ref{eq:measurement2})
as a proper measurement is questionable. However, this shortcomming is
solvable ---at least from a theoretical point of view--- just by
considering that the apparatus represents, not only the measuring
machine, but also the memory of the observer. We will rely on this
interpretation throughout the rest of the paper; technical
considerations are far beyond its scope. Therefore, from now on, the
state of any measuring apparatus will represent the memory record of
any quantum machine playing the observer role. After the complete
protocol is finished, all these records are supposed to be available
as the outputs of the quantum computation.

The second one is the basis ambiguity problem \cite{Zurek:03}. The
very same state in Eq. (\ref{eq:measurement2}), $\ket{\Psi_1}$, can be
written in different basis,
\begin{equation}
  \ket{\Psi_1} = \frac{1}{\sqrt{2}} \left( \ket{\alpha} \ket{A_{\alpha}} + \ket{\beta} \ket{A_{\beta}} \right),
\label{eq:basis}  
\end{equation}
where
\begin{subequations}
  \begin{align}
  \ket{\alpha} &= \sin \theta \ket{h} + \cos \theta \ket{v}, \\
  \ket{\beta} &= - \cos \theta \ket{h} + \sin \theta \ket{v}, \\
  \ket{A_{\alpha}} &= \sin \theta \ket{A_h} + \cos \theta \ket{A_v}, \\
  \ket{A_{\beta}} &= - \cos \theta \ket{A_h} + \sin \theta \ket{A_v}.
  \end{align}
\end{subequations}
That is, the final state of the very same measuring protocol, starting
from the very same initial condition, can also be written as the
superposition given in Eq. (\ref{eq:basis}) for arbitrary values of
$\theta$. This problem blurs the usual interpretation of all the
versions of the Wigner's friend experiment. {\em As Eq. \eqref{eq:basis} is a correct representation of agent's $I$ memory, we have no grounds to conclude that 
outcome of its measurement is either $h$ or $v$, instead of $\alpha$
or $\beta$}. The unitary evolution giving rise to the measurement,
Eq. (\ref{eq:measurement1}), does not determine a preferred basis for
the corresponding definite outcome. Hence, a physical mechanism for the emergence of such an outcome must be provided, in order to not get stuck on a fuzzy interpretation issue. The main trademark of the decoherence formalism is providing a plausible mechanism. 

There are several ways to solve this problem. One of them consists in
modifying the Schr\"odinger equation to model the wavefunction
collapse and to choose the corresponding preferred basis. These
theories are based on the fact that superpositions have been
experimentally observed in systems up to $10^{-21}$ g, whereas the
lower bound for a classical apparatus is around $10^{-6}$ g \cite{Bassi:13}. This
means that the Schr\"odinger equation is just an approximation, which
works pretty well for small systems, but fails for systems as large as
measurement devices. Such a real collapse would change all the
dynamics of Wigner's friend experiments, presumably ruling out all
their inconsistencies.

Another possibility, the one which is the object of this work, is that
Eq. (\ref{eq:measurement1}) is {\em not} a complete measurement, but
just a pre-measurement ---a previous step required for any
observation \cite{Zurek:03,Zurek:07}. Following this interpretation,
the observation is not completed until a third party, an environment
which is {\em not} the object of the measurement, becomes correlated
with the measured system and the measuring apparatus. This correlation
is given again by a Hamiltonian, and therefore consists in a unitary
evolution. If such an environment is continously monitorizing the
system \cite{nota0}, the state of the whole system becomes
\begin{equation}
\ket{\Psi_2} = \frac{1}{\sqrt{2}} \left( \ket{h} \ket{A_{h}} \ket{\varepsilon_1 (t)}  + \ket{v} \ket{A_{v}} \ket{\varepsilon_2 (t)} \right),
\label{eq:environment1}
\end{equation}
where the states of the environment $\ket{\varepsilon_1(t)}$ and
$\ket{\varepsilon_2(t)}$ change over time, because the apparatus
is continuously interacting with it, and $\braket{\varepsilon_1(t)}{\varepsilon_1(t)} = \braket{\varepsilon_2(t)}{\varepsilon_2(t)} = 1$. Note that
Eq. (\ref{eq:environment1}) entails that the correlations betweem the
system and the apparatus remain untouched despite the continuous
monitorization by the environment. Hence, the states $\ket{A_h}$ and
$\ket{A_v}$ are called {\em pointer states}, because they represent
the stable states of the apparatus \cite{Zurek:03,Zurek:81} and the stable records in the memory of the observers. Furthermore, if
such an apparatus-environment interaction implies
$\braket{\varepsilon_1(t)}{\varepsilon_2 (t)}=0$, $\forall t>\tau$,
where $\tau$ can be understood as the time required to complete the
measurement, the following affirmations hold:

{\em (i)} There is no other triorthogonal basis to write the state
given by Eq. (\ref{eq:environment1}) \cite{Elby:94}. That is, the
basis ambiguity problem is fixed by the action of the uncontrolled
environment.

{\em (ii)} As the observer $I$ cannot measure the environment, its
memory record and all the further experiments it can perform on the
system and the apparatus are compatible with the following mixed state
\begin{equation}
  \rho = \frac{1}{2} \left( \ket{h} \ket{A_h} \bra{h} \bra{A_h} + \ket{v} \ket{A_v} \bra{v} \bra{A_v} \right),
  \label{eq:colapso}
\end{equation}
independently of the particular shapes of both $\ket{\varepsilon_1
  (t)}$ and $\ket{\varepsilon_2 (t)}$. That is, the observer $I$ sees
the system as if it were randomly collapsed either to $\ket{h}
\ket{A_h}$ or to $\ket{v} \ket{A_v}$, even though the real evolution
of the complete system, {\em including itself!}, is deterministic and
given by Eq. (\ref{eq:environment1}). Relying on the decoherence
framework, such an observer can only deduce that the real state of the
system, the apparatus and itself must be something like
Eq. (\ref{eq:environment1}) \cite{nuevanota}. Randomness arises
through this lack of knowledge.

At this point, it is worth remarking that the decoherent environment
must be understood as a fundamental part of the measuring device, not
a practical difficulty under realistic condictions ---the difficulty
of keeping the system aside from external perturbations. If the
decoherence framework is applied, any quantum machine must include
such an environment as an inseparable part of it. The trademark of
this framework is postulating that definite ---classical--- outcomes
arise as a consequence of the continuous environmental monitorization;
if such an environment does not exist, no definite outcomes are
observed. In other words, the observation is completed when the state
given by Eq. (\ref{eq:environment1}) is reached: {\em if the observer
  sees a collapsed state is because an uncontrolled environment is
  monitorizing the system (including itself!), and thus the complete
  wavefunction is given by Eq. (\ref{eq:environment1})}. The
decoherence interpretation of quantum measurements also provides a
framework to derive the Born rule from fundamental postulates
\cite{Zurek:07}. Notwidthstanding, all this work is based just on the
previous facts {\em (i)} and {\em (ii)}, and therefore the possible
issues in this derivation of the Born rule are not relevant.

Before ending this section, it is interesting to delve into the
differences between the standard interpretation of quantum
measurements and the one supplied by the decoherence formalism. Under
normal circumstances, both interpretations provide indistinguishable
results. For example, the standard interpretation establises that,
once an agent has observed a definite outcome in a polarisation
experiment, say $h$, then any further measurements performed in the
same basis are bounded to give the same outcome, $h$. This important fact is exactly reproduced by the decoherence framework. A second measurement with an identical apparatus, denoted $A'$, performed on Eq. \eqref{eq:environment1} will give
\begin{equation}
\ket{\Psi_3} = \coef{1}{2} \left( \ket{h} \ket{A_h} \ket{A'_h} \ket{\epsilon_1(t)} + \ket{v} \ket{A_v} \ket{A'_v} \ket{\epsilon_2(t)} \right),
\end{equation}
if we logically assume that the Hamiltonian modelling the interaction between the apparatus and the environment is identical for two identical apparati. Therefore, the perception of the observer is given by
\begin{equation}
  \rho = \frac{1}{2} \left( \ket{h} \ket{A_h} \ket{A'_h} \bra{h} \bra{A_h} \bra{A'_h} + \ket{v} \ket{A_v} \ket{A'_v} \bra{v} \bra{A_v} \bra{A'_v} \right).
\end{equation}
That is, its internal memory says that if it has observed $h$ in the first measurement, then it has also observed $h$ in the second.

In the next sections we will show that the standard interpretation and
the decoherence formalism do show important differences when the
observers are the object of external interference experiments. The key
point lies, again, in the role played by the environment. To perform a
proper interference experiment, the external observer must act
coherently on the system, the apparatus (that is, the memory of the
internal agent), {\em and} the environment. As a consequence of this
action, the state of the environment will eventually change in a
perfectly predictable way. And, as it is the ultimate responsible of
the definite outcome observed by the internal agent, its internal
memory will also change accordingly. We will discuss below how these
changes release quantum theory from inconsistencies.

\subsection{A simple model for the laboratories}
\label{sec:model}

The laboratories in which agents $I$ perform their measurements are
quantum machines evolving unitarily. Their Hamiltonians must consist
of: {\em (i)} a system-apparatus interaction, performing the
pre-measurements; and {\em (ii)} an apparatus-environment interaction,
following the decoherence formalism. For (i) we consider the logical
C-NOT gate given in Eq. (\ref{eq:medida}). Following \cite{Zurek:03},
for (ii) we propose a model
\begin{equation}
  \begin{split}
    H &=\ket{A_h}\bra{A_h} \sum_{n,m} V^h_{nm} \ket{\varepsilon_n} \bra{\varepsilon_m} + \\ &+ \ket{A_v}\bra{A_v} \sum_{n,m} V^v_{nm} \ket{\varepsilon_n} \bra{\varepsilon_m},
    \end{split}
  \label{eq:interaccion}
\end{equation}
where $V^h$ and $V^v$ are the coupling matrices giving rise to the
interaction. The only condition for them is to be hermitian matrices;
independently of their particular shapes, the Hamiltonian given by
Eq. (\ref{eq:interaccion}) guarantees that the correlations $\ket{h}
\ket{A_h}$ and $\ket{v} \ket{A_v}$ remain unperturbed, that is,
$\ket{A_h}$ and $\ket{A_v}$ are the pointer states resulting from this
interaction, and the state given by Eq. (\ref{eq:environment1}) holds
for any time.

To build a simple model, we consider that both $V^h$ and $V^v$ are
real symmetric random matrices of the Gaussian Orthogonal Ensemble
(GOE), which is the paradigmatic model for quantum chaos
\cite{Gomez:11}. They are symmetric square matrices of size $N$, with
independent Gaussian random elements with mean $\mu(V_{nm})=0$,
$\forall n,m=1, \ldots, N$, and standard deviation $\sigma(V_{nn})=1$, $\forall
n=1, \ldots, N$ (diagonal elements); and $\sigma(V_{nm})=1/\sqrt{2}$,
$\forall n\neq m=1, \ldots, N$ (non-diagonal elements).

\begin{figure}
  \begin{tabular}{c}
    \includegraphics[width=0.5\columnwidth]{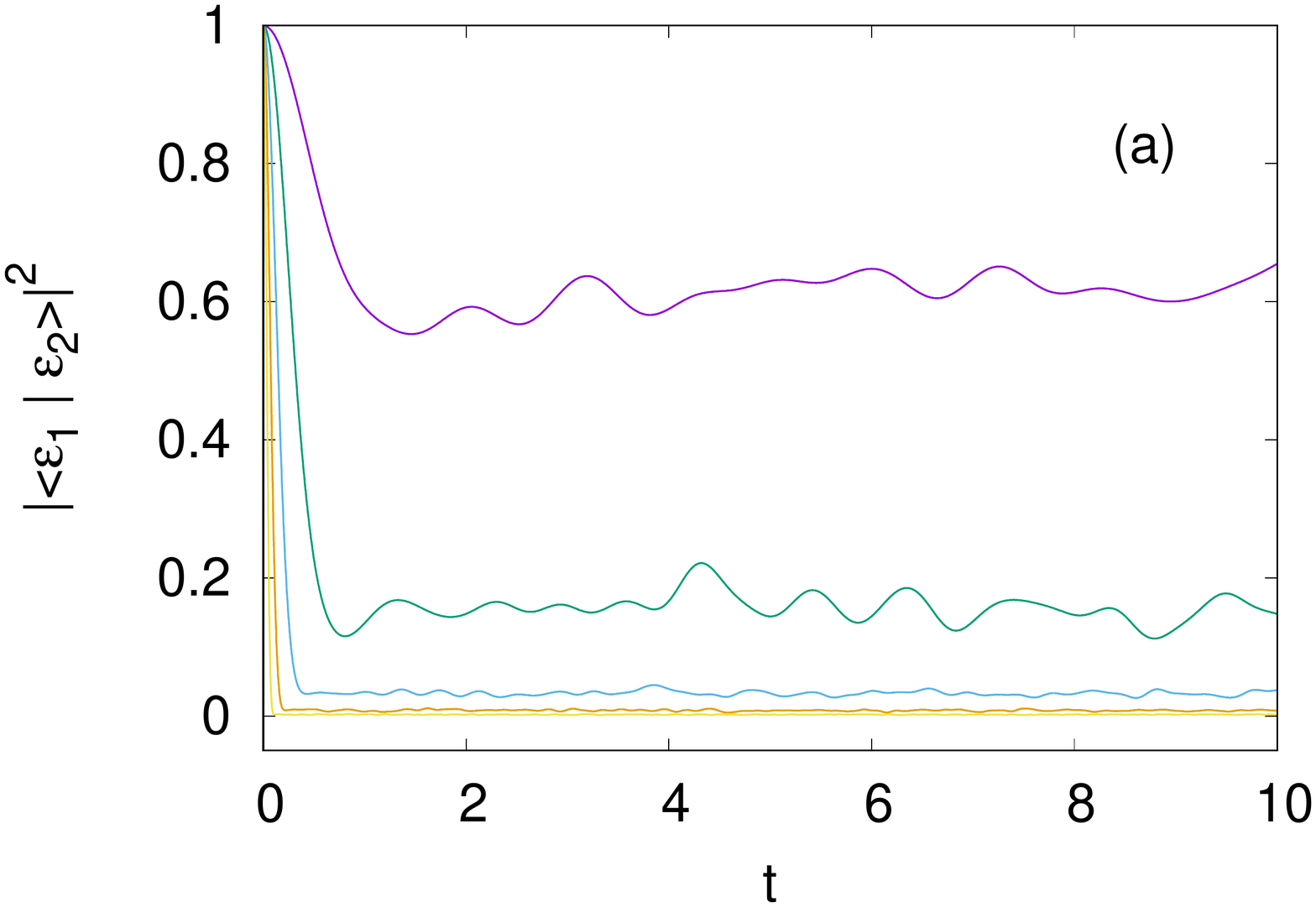} \\[-1cm]
    \includegraphics[width=0.5\columnwidth]{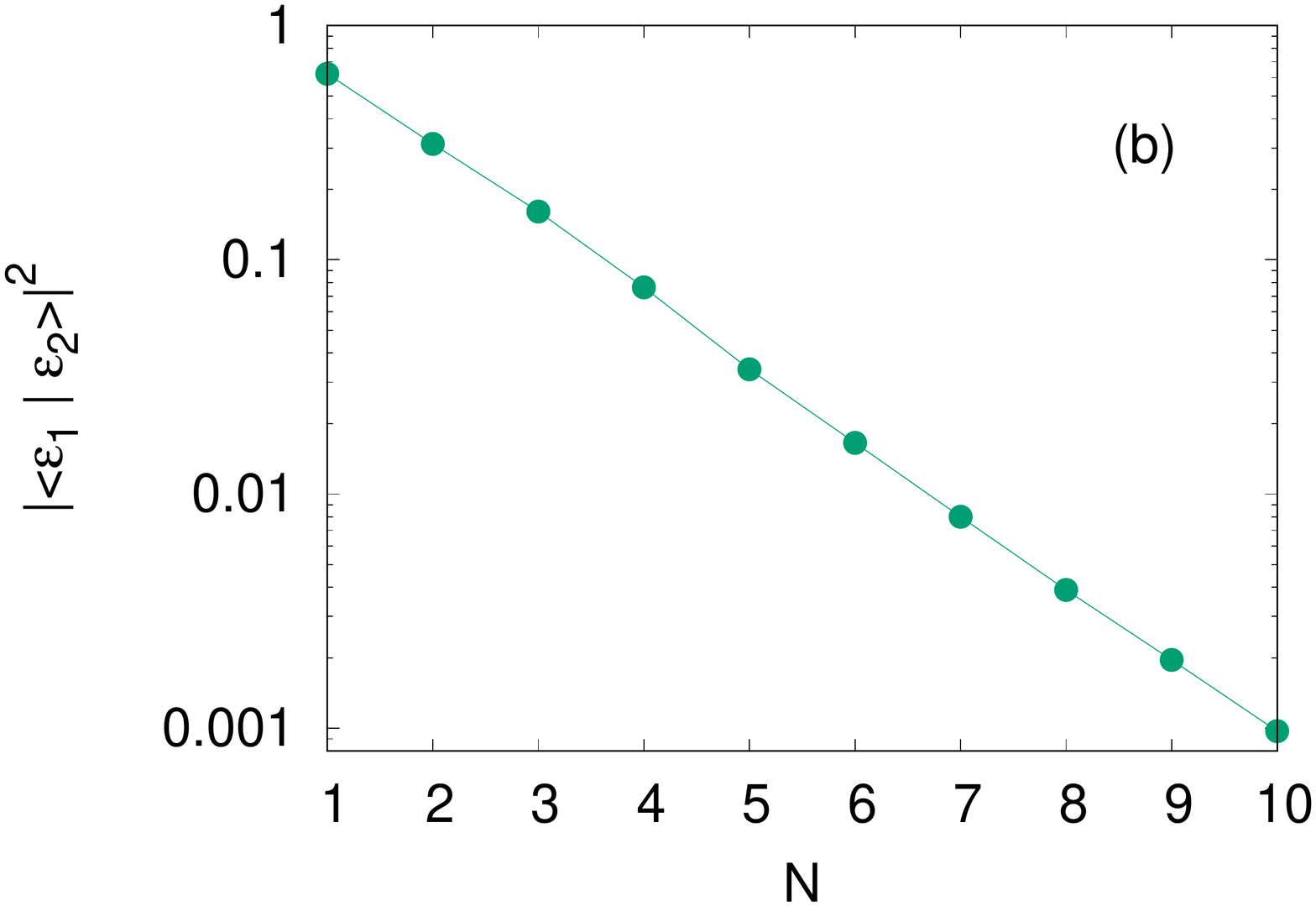}
    \end{tabular}
  \caption{Panel (a), value of $\left|
    \braket{\varepsilon_1(t)}{\varepsilon_2(t)} \right|^2$ as a
    function of time, for environments composed by different number of
    qbits. The solid curves show, from the upper one to the lower one,
    $N=1$, $N=3$, $N=5$, $N=7$ and $N=9$. Panel (b),
    finite-size scaling for the long-time average of $\left|
    \braket{\varepsilon_1(t)}{\varepsilon_2(t)} \right|^2$, as a
    function of the number of qbits composing the environment, $N$.}
  \label{fig:ortonormal}
\end{figure}

In panel (a) of Fig. \ref{fig:ortonormal} we show how the overlap
between the two states of the environment, $\ket{\varepsilon_1(t)}$
and $\ket{\varepsilon_2(t)}$, evolves with time; in panel (b) how it
evolves with the environment size. To perform the
calculations, we have considered that the environment consists in $N$
qbits, and hence the dimension of its Hilbert space is $d=2^N$. In all
the cases, the initial state is a tensor product
\begin{equation}
\ket{\Psi(0)} = \frac{1}{\sqrt{2}} \left[ \ket{h} \ket{A_h} + \ket{v} \ket{A_v} \right] \otimes \ket{\epsilon_0},
\end{equation}
where $\ket{\epsilon_0}$ is the first element of the environmental
basis (as the interaction is a GOE random matrix, the particular shape
of the basis is irrelevant \cite{Gomez:11}). All the results are
averaged over $50$ different realizations. We have considered
$\hbar=1$.

Panel (a) of Fig. \ref{fig:ortonormal} shows the results of $\left|
\braket{\varepsilon_1(t)}{\varepsilon_2(t)} \right|^2$ for $N=1$
($d=2$), $N=3$ ($d=8$), $N=5$ ($d=32$), $N=7$ ($d=128$), and $N=9$
($d=512$). We clearly see that, the larger the number of environmental
qbits, the smaller the value of $\left|
\braket{\varepsilon_1(t)}{\varepsilon_2(t)} \right|^2$ at large times,
and the smaller the characteristic time $\tau$ required to complete
the measurement process. Therefore, the condition $\left|
\braket{\varepsilon_1(t)}{\varepsilon_2(t)} \right|^2 \sim 0$ is fast
reached if the number of the environmental qbits is $N \sim 10$. The
results plotted in panel (b) of the same figure confirm this
conclusion. We show there the long-time average of $\left|
\braket{\varepsilon_1(t)}{\varepsilon_2(t)} \right|^2$, calculated for
$2 \leq t \leq 10$, as a function of the number of environmental
qbits. It is clearly seen that the overlap between these states
decreases fast with this number. As a consequence, we can safely
conclude that an agent $I$ operating within a laboratory described by
Eq. (\ref{eq:interaccion}) will observe a state given by
Eq. (\ref{eq:colapso}).

\begin{table}
  \begin{tabular}{p{0.5cm}p{7.5cm}}
    \hline
{\bf L1} & The measured system. \\

{\bf L2} & The measuring apparatus. \\

{\bf L3} & An internal environment, with a chaotic interaction like
the one given by Eq. (\ref{eq:interaccion}), and large enough to
guarantee $\left| \braket{\varepsilon_1(t)}{\varepsilon_2(t)}
\right|^2 \sim 0$. \\\hline

  \end{tabular}
  \caption{Parts of laboratories in which the agents $I$ perform their measurements in a Wigner's friend experiment, following the decoherence framework.}
\label{tab:laboratorio}
\end{table}

These results imply that the laboratories in which all the agents
perform their measurements must have the structure summarized in
Tab. \ref{tab:laboratorio}. It is worth to note that this structure is
independent from any further evolution of the measured system, after
the pre-measurement is completed. For example, let us imagine that the
measured system has its own Hamiltonian, and therfore the time
evolution for the whole system is governed by
\begin{equation}
H = H_S \otimes I_{A \varepsilon} + I_S \otimes H_{A \varepsilon},
\end{equation}
where $H_S$ is the Hamiltonian for the measured system, $H_{A
  \varepsilon}$ represents the environment-apparatus interaction,
given by Eq. (\ref{eq:interaccion}), and $I_S$ ($I_{A \varepsilon}$)
is the identity operator for the system (environment-apparatus). As
the two terms in this Hamiltonian commute pairwise, the time
evolution of the whole system is
\begin{equation}
  \ket{\Psi(t)} = \coef{1}{2} \ket{\eta(t)}\ket{A_h}\ket{\varepsilon_1 (t)} + \coef{1}{2} \ket{\upsilon(t)} \ket{A_v} \ket{\varepsilon_2(t)},
\end{equation}
where the notation $\eta(t)$ and $\upsilon(t)$ has been
chosen to denote that $\eta(t)$ is the state which evolves from an
initial condition consisting in an horizontally polarised photon,
$\ket{\eta(t)} = \exp\left(-i H_S t \right) \ket{h}$, and $\upsilon(t)$ the
state which evolves from a vertically polarised photon, $\ket{\upsilon(t)} =
\exp\left(-i H_S t \right) \ket{v}$. Therefore all further
measurements of the same agents are well described by
\begin{equation}
  \rho(t) = \frac{1}{2} \ket{\eta(t)} \ket{A_h} \bra{\eta(t)} \bra{A_h} + \ket{\upsilon(t)} \ket{A_v} \bra{\upsilon(t)} \bra{A_v}.
 \label{eq:memoria}
  \end{equation}
That is, all the possible experiments that agent $I$ can perform in
the future are compatible with the system collapsing onto either
$\ket{h}$ or $\ket{v}$ after the measurement, and unitarily evolving
from the corresponding initial condition. In other words, and as we
have already pointed out, this framework is fully compatible with the
Copenhaguen interpretation$\ldots$ {\em but the wave-function collapse
  being just a consequence of ignoring the environmental degrees of
  freedom.}  It is worth to remark that this is not a subjective
interpretation, but the result of a unitary time evolution including a
number of degrees of freedom that cannot be measured by the same
observer. Eq. \eqref{eq:memoria} establishes that a further reading of
the agent memory record would reveal that the photon has collapsed
either to $h$ or $v$, and then it has evolved from the corresponding
initial condition.

\begin{figure}
  \begin{tabular}{c}
    \includegraphics[width=0.5\columnwidth]{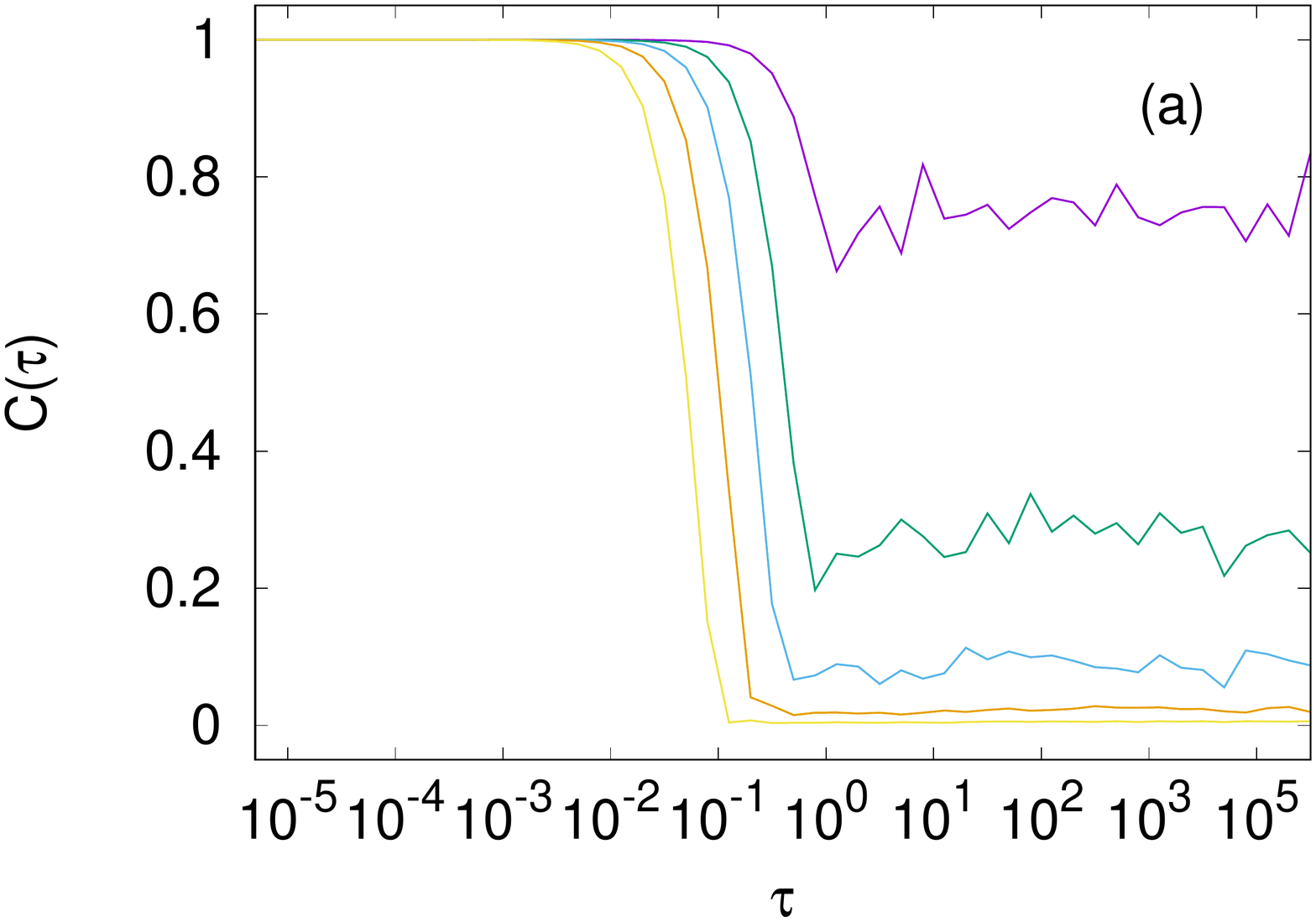} \\[-1cm]
    \includegraphics[width=0.5\columnwidth]{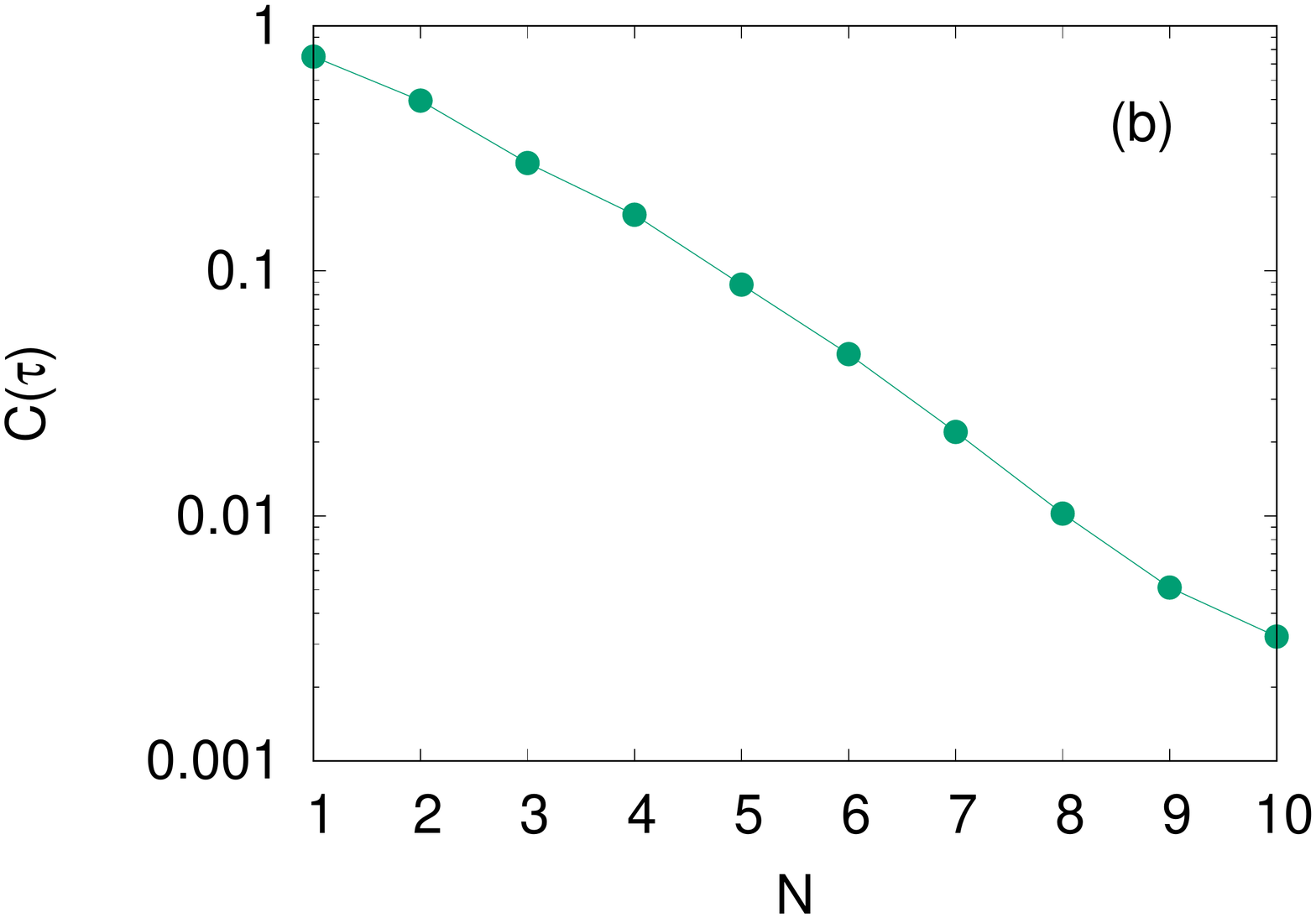}
    \end{tabular}
  \caption{Panel (a), value of $C(\tau)$ as a
    function of $\tau$, for environments composed by different number of
    qbits. The solid curves show, from the upper one to the lower one,
    $N=1$, $N=3$, $N=5$, $N=7$ and $N=9$. Panel (b),
    finite-size scaling for the long-time average of $C(\tau)$, as a
    function of the number of qbits composing the environment, $N$.}
    \label{fig:correlacion}
\end{figure}

As the key point in Wigner's friend experiments consists in further
interference measurements on the whole laboratory, a
study of the complexity of the state resulting from the time evolution
summarized in Fig. \ref{fig:ortonormal} is necessary. Such a study can
be made by means of a correlation function $C(\tau)=\left|
\braket{\varepsilon_1(t)}{\varepsilon_1(t+\tau)} \right|^2$. If
$C(\tau) \sim 1$, then the time evolution of the environmental state
$\ket{\varepsilon_1(t)}$ is quite simple; its only possible change is
an irrelevant global phase. Such a simple evolution would facilitate 
further interference experiments. On the contrary, if $C(\tau)$ quickly
decays to zero, the same evolution is highly involved, implying that the
state of the whole laboratory is complex enough to hinder
further interference experiments.

Results are summarized in Fig. \ref{fig:correlacion}. Panel (a) shows
$C(\tau)$ for the same environments displayed in the same panel of
Fig. \ref{fig:ortonormal}. It has been obtained after a double
average: over $50$ different realizations, and over $10^4$ different
values of the time $t$. Panel (b) of Fig. \ref{fig:correlacion}
displays a finite size scaling of $C(\tau)$ for large values of time
versus the number of environmental qbits, calculated averaging over $\tau \geq 10$. It is clearly seen that the
results shown in this Figure are correlated with the ones displayed in
Fig. \ref{fig:ortonormal}. That is, if the environment is large enough
to give rise to $\left| \braket{\varepsilon_1(t)}{\varepsilon_2(t)}
\right|^2 \sim 0$, then the environmental states fulfill $C(\tau) \sim
0$; the smaller the overlap between $\ket{\varepsilon_1(t)}$ and
$\ket{\varepsilon_2(t)}$, the smaller the value of the correlation
function $C(\tau)$. It is also worth noting that $C(\tau)$ decays
very fast to zero; for $N=9$, $C(\tau) \sim 0$ for $\tau \simeq
10^{-1}$. This means that the state of the environment is changing
fast, and therefore the state of the whole laboratory, including the
measured system, the measuring apparatus and the environment, is very
complex.

As we have pointed out above, the key point of all the versions of
Wigner's friend experiments consist in further interference
measurements performed by an external agent, for which the whole
laboratory evolves unitarily following
Eq. (\ref{eq:environment1}). Both in its original \cite{Wigner:61} and
its extended versions, discussed in
\cite{Renner:18,Bruckner:18,Proietti:19}, the external agents perform
interference experiments involving only two states, $\ket{A_h}
\ket{h}$ and $\ket{A_v} \ket{v}$. The Hilbert spaces of the simplified
versions of the laboratories discussed in these papers are spanned by
$\left\{ \ket{A_h} \ket{h}, \ket{A_v} \ket{v}, \ket{A_h} \ket{v},
\ket{A_v} \ket{h} \right\}$. Notwithstanding, the last two states are
never occupied, and hence such two-state interference experiments are
feasible \cite{Proietti:19}. The situation arising from the
decoherence framework is far more complex. The dimension of the whole
laboratory, composed by the measured system, the measuring apparatus,
and an environment with $N$ qbits, is $d=2^{N+2}$. From the results
summarized in Fig. \ref{fig:correlacion}, we conjecture that all the
$2^N$ states of the environment are populated, and therefore $2^{N+1}$
states of the whole laboratory become relevant for further
interference experiments. Hence, the first consequence of the results
discussed in this section is that experiments like the ones in
\cite{Wigner:61,Renner:18,Bruckner:18,Proietti:19} become extremely
difficult. However, as $\left|
\braket{\varepsilon_1(t)}{\varepsilon_2(t)} \right|^2 \sim 0$, it is
true that only two states, $\ket{h} \ket{A_h} \ket{\varepsilon_1(t)}$
and $\ket{v} \ket{A_v} \ket{\varepsilon_2(t)}$, are populated {\em at
  each time $t$}; the rest of the Hilbert space is irrelevant {\em at
  that particular value of the time $t$}. Unfortunately, these states
change very fast with time, and in a very complex way. Therefore, an
interference experiment involving only two states, $\ket{h} \ket{A_h}
\ket{\varepsilon_1(t)}$ and $\ket{v} \ket{A_v}
\ket{\varepsilon_2(t)}$, would require a very restrictive protocol,
whose main requisites are summarized in Tab. \ref{tab:protocol}. Only
if such requisites are fulfilled, the external agent $E$ can rely on a
simplified basis, composed by $\ket{h(\tau)} \equiv \ket{h} \ket{A_h}
\ket{\varepsilon_1(\tau)}$ and $\ket{v(\tau)} \equiv \ket{v} \ket{A_v}
\ket{\varepsilon_2(\tau)}$, where $\tau=t_E - t_I$, $t_I$ the time at
which agent $I$ performs its measurement, and $t_E$ the same for agent
$E$.  A small error in points R1-R4 would imply that the real state of
the laboratory, $\ket{\Psi(t)}$, had negligible overlaps with both
$\ket{h(\tau)}$ and $\ket{v(\tau)}$, and therefore any interference
experiments involving just these two states would give no
significative outcomes. Notwithstanding, given the promising
state-of-the-art in quantum computing \cite{google}, we can trust in
future quantum machines able to work with enough precission.

\begin{table}
  \begin{tabular}{p{0.5cm}p{7.5cm}}
    \hline {\bf R1} & A perfect knowlede of the interaction between
    the system and the apparatus, $H$, given by
    Eq. (\ref{eq:interaccion}). \\

{\bf R2} & A perfect knowledge of the environmental initial state,
$\ket{\varepsilon_0}$. \\

{\bf R3} & A perfect knowledge of the time at which agent $I$ performs its measurement, $t_I$. \\

{\bf R4} & A perfect choice of the time at which agent $E$ performs
its interference experiment, $t_E$. \\\hline
  \end{tabular}
  \caption{Requisites for an extended Wigner's friend experiment in which the external agent, $E$, performs an interference experiment involving only two states.}
  \label{tab:protocol}
  \end{table}

Before applying these conclusions to the original and the extended
versions of the Wigner's friend experiments, it makes sense to test if
these conclusions depend on the particular model we have chosen for
the apparatus-environment interaction. To tackle this task, we
consider more general random matrices $V^h$ and $V^v$ in
Eq. (\ref{eq:interaccion}), in which $\mu(V_{nm})=0$, $\forall n,m=1,
\ldots, N$, $\sigma(V_{nn})=1$, $\forall n=1, \ldots, N$ (diagonal
elements); and $\sigma(V_{nm}) = 1/\left( \sqrt{2} \left| n-m
\right|^{\alpha}\right)$ $\forall n \neq m=1, \ldots N$ (non-diagonal
elements). If the parameter $\alpha$ is large, then only very few
non-diagonal elements are relevant, and hence the interaction becomes
approximately integrable. On the contrary, if $\alpha=0$, GOE
(chaotic) results are recovered.

We fix our attention in the degree of chaos of the resulting
Hamiltonian. To do so, we study the ratio of consecutive level
spacings distribution, $P(r)$, where $r_n=s_{n+1}/s_{n}$ and
$s_n=E_{n+1} - E_n$, $\left\{ E_n \right\}$ being the energy spectrum
of the system. It has been shown \cite{Atas:13} that the distribution
for standard integrable systems is $P(r)=1/(1+r)^2$, whereas it is
$P(r)=27(r+r^2)/\left(8 (1+r+r^2)^{5/2} \right)$ for GOE systems; a generic interpolating distribution has been recently proposed \cite{Corps:19}.

\begin{figure}
  \begin{tabular}{cc}
    \includegraphics[width=0.5\columnwidth]{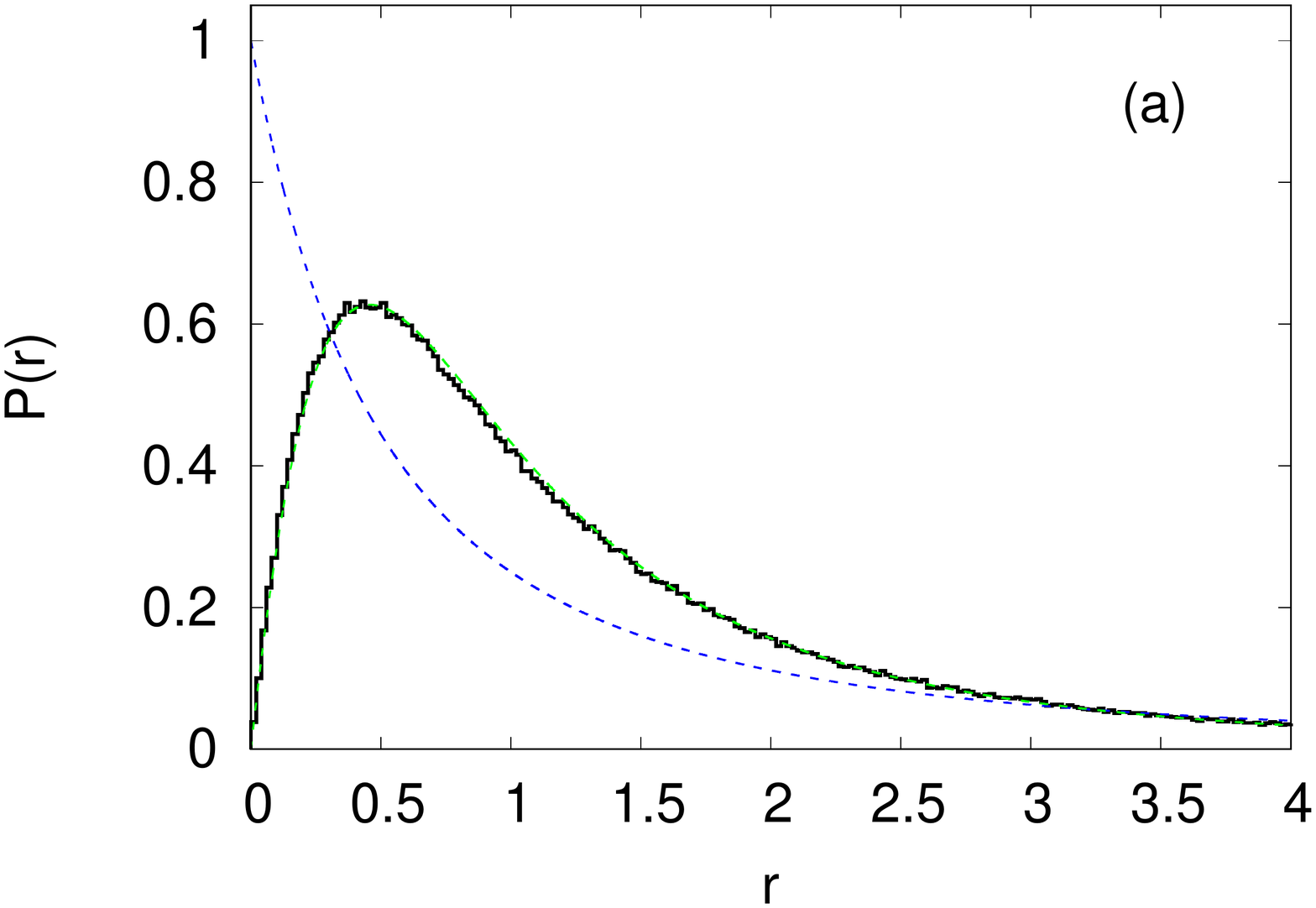} &
    \includegraphics[width=0.5\columnwidth]{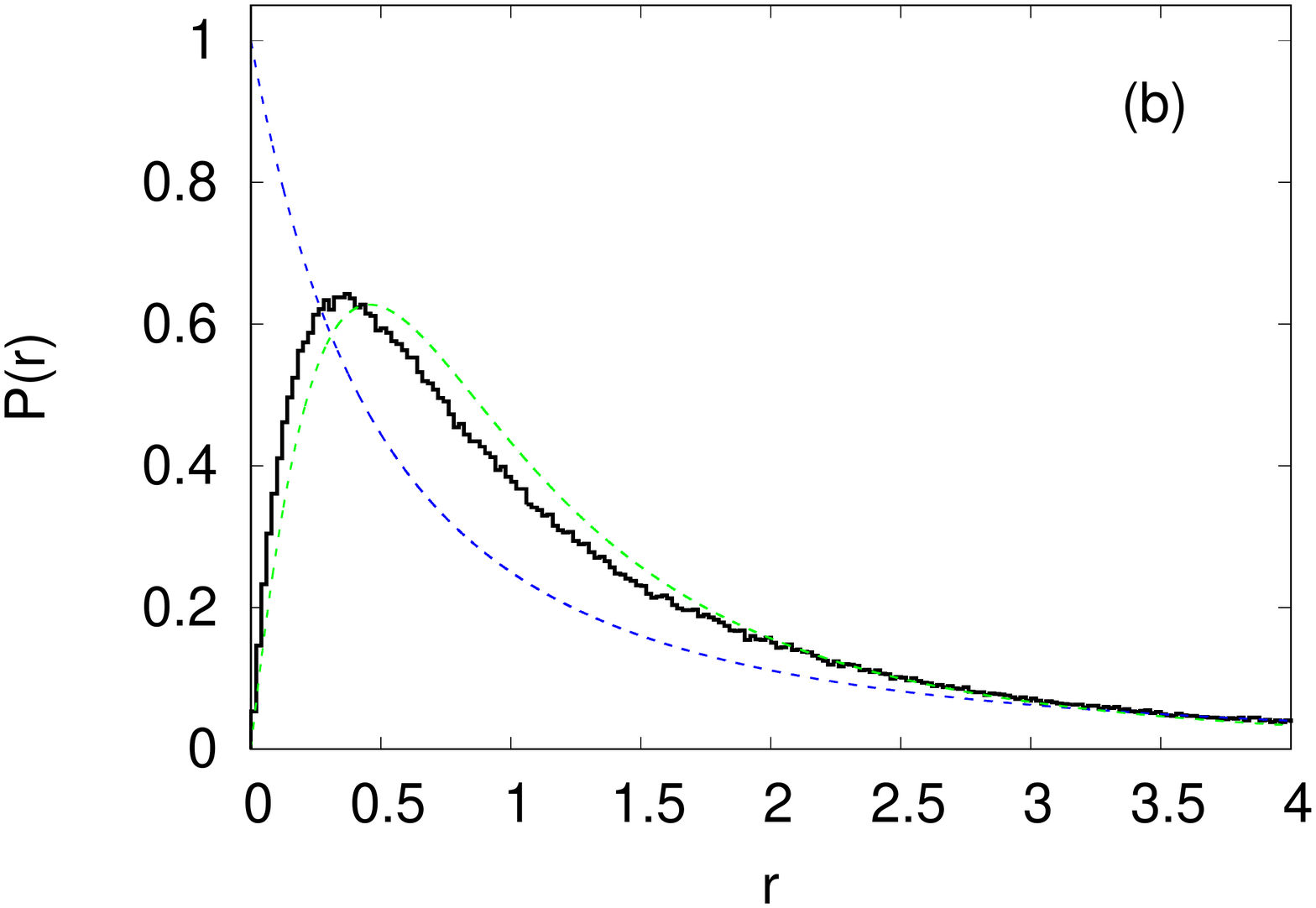} \\[-0.8cm]
    \includegraphics[width=0.5\columnwidth]{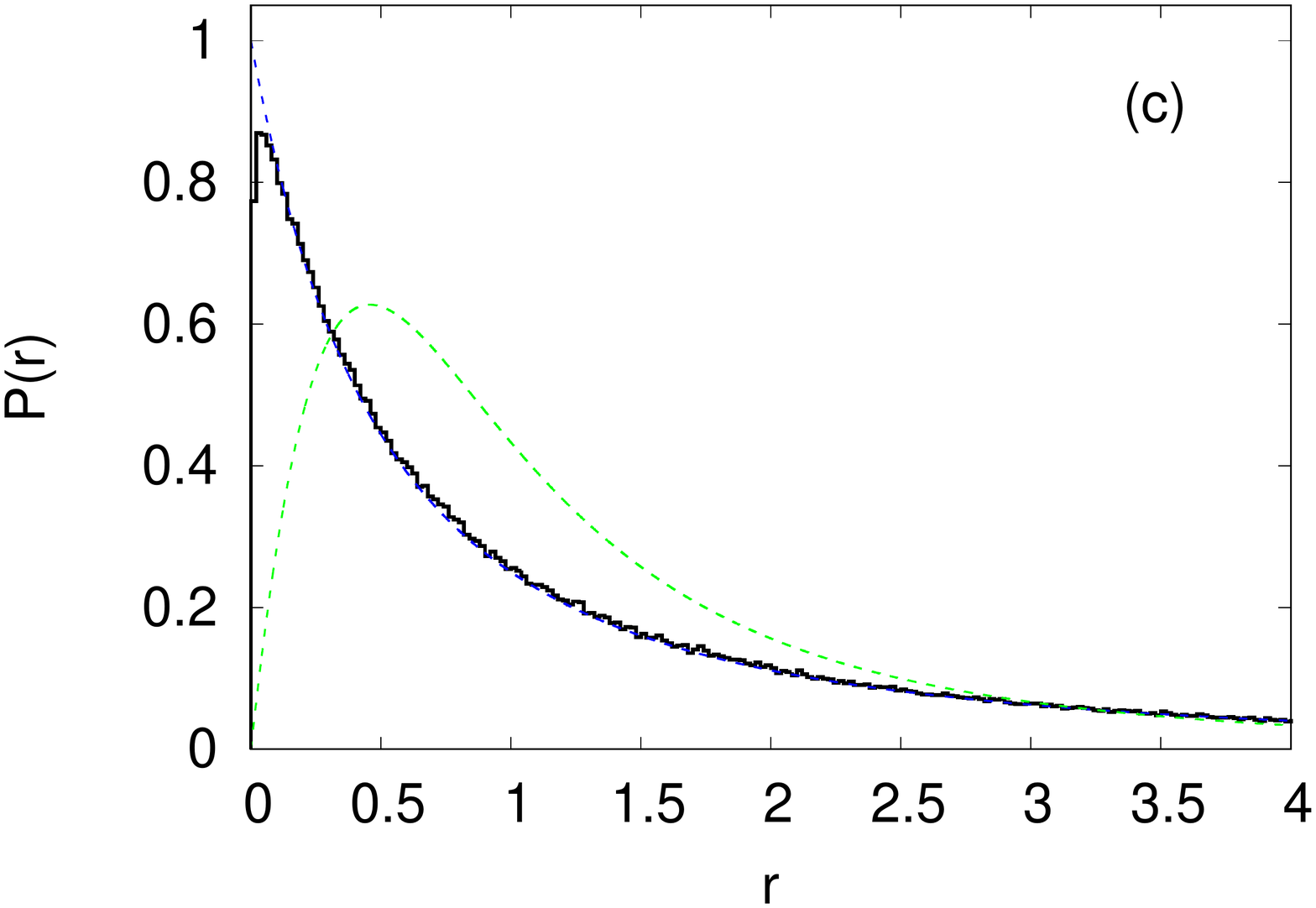} &
    \includegraphics[width=0.5\columnwidth]{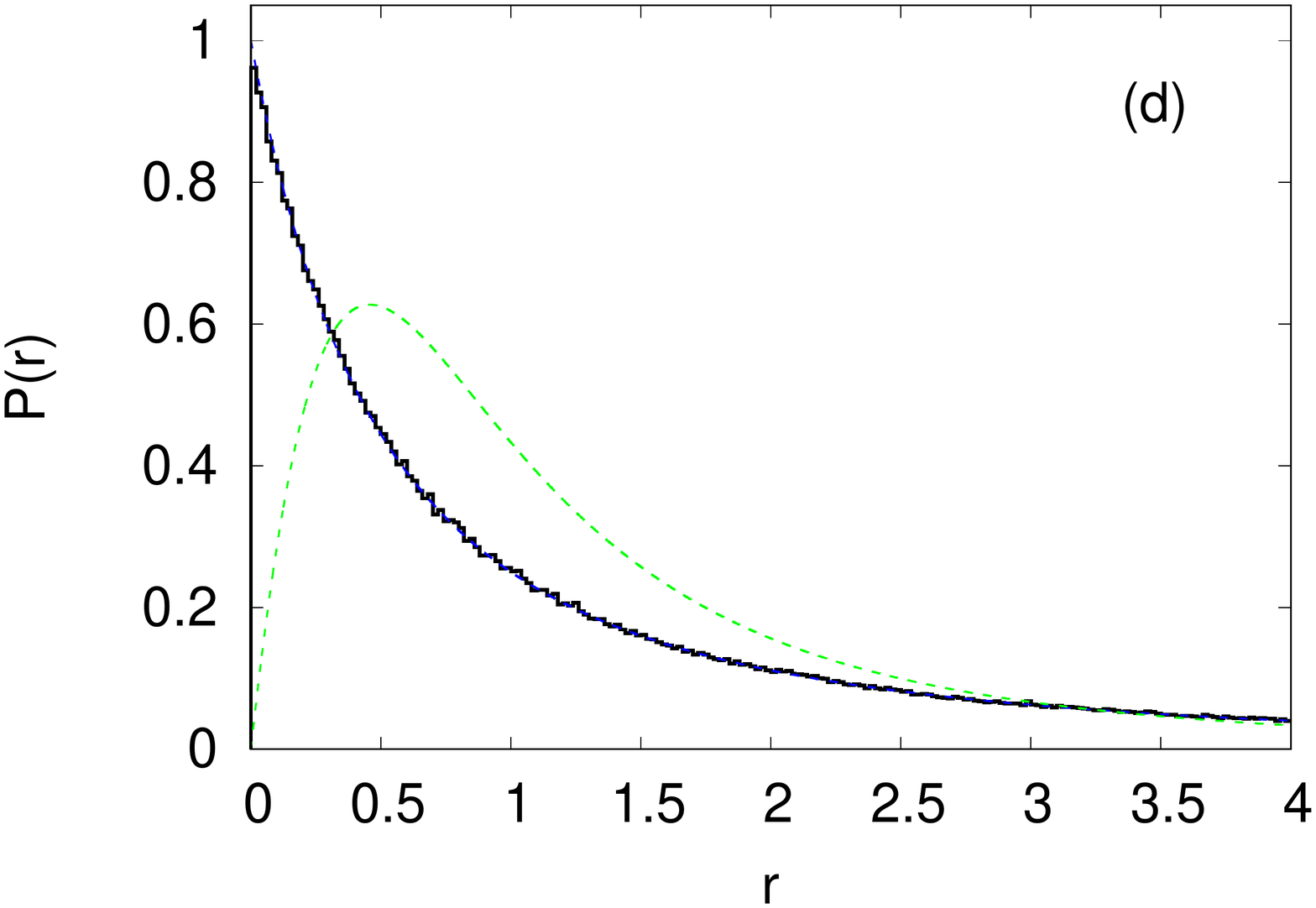} 
    \end{tabular}
  \caption{Ratio of consecutive level spacings distribution, $P(r)$, for $\alpha=0.5$ [panel (a)], $\alpha=1$ [panel (b)], $\alpha=2$ [panel (c)], and $\alpha=4$ [panel (d)]. Solid histograms show the numerical results for $2000$ matrices with dimension $d=512$; green dashed line, the result for a GOE system, $P(r)=27(r+r^2)/\left(8 (1+r+r^2)^{5/2} \right)$, and the blue dashed line, the result for an integrable system, $P(r)=1/(1+r)^2$.}
  \label{fig:estadistica}
\end{figure}

In Fig. \ref{fig:estadistica} we show the results for four different
values of $\alpha$, $\alpha=0.5$, $\alpha=1$, $\alpha=2$,
$\alpha=4$. They consist in the average over $2000$ realizations of
matrices of dimension $d=512$. The case with
$\alpha=0$ (not shown) exactly recovers the GOE result, as
expected. The case with $\alpha=0.5$ [panel (a)] is also fully
chaotic; its ratio of consecutive level spacings distribution, $P(r)$, is identical to
the GOE result. Things become different for larger values of
$\alpha$. The case $\alpha=1$ [panel (b)] is yet different from the
GOE result, althought its behavior is still highly chaotic. The cases
$\alpha=2$ [panel (c)] and $\alpha=4$ [panel (d)] are very close to
the integrable result.

\begin{figure}
    \includegraphics[width=0.5\columnwidth]{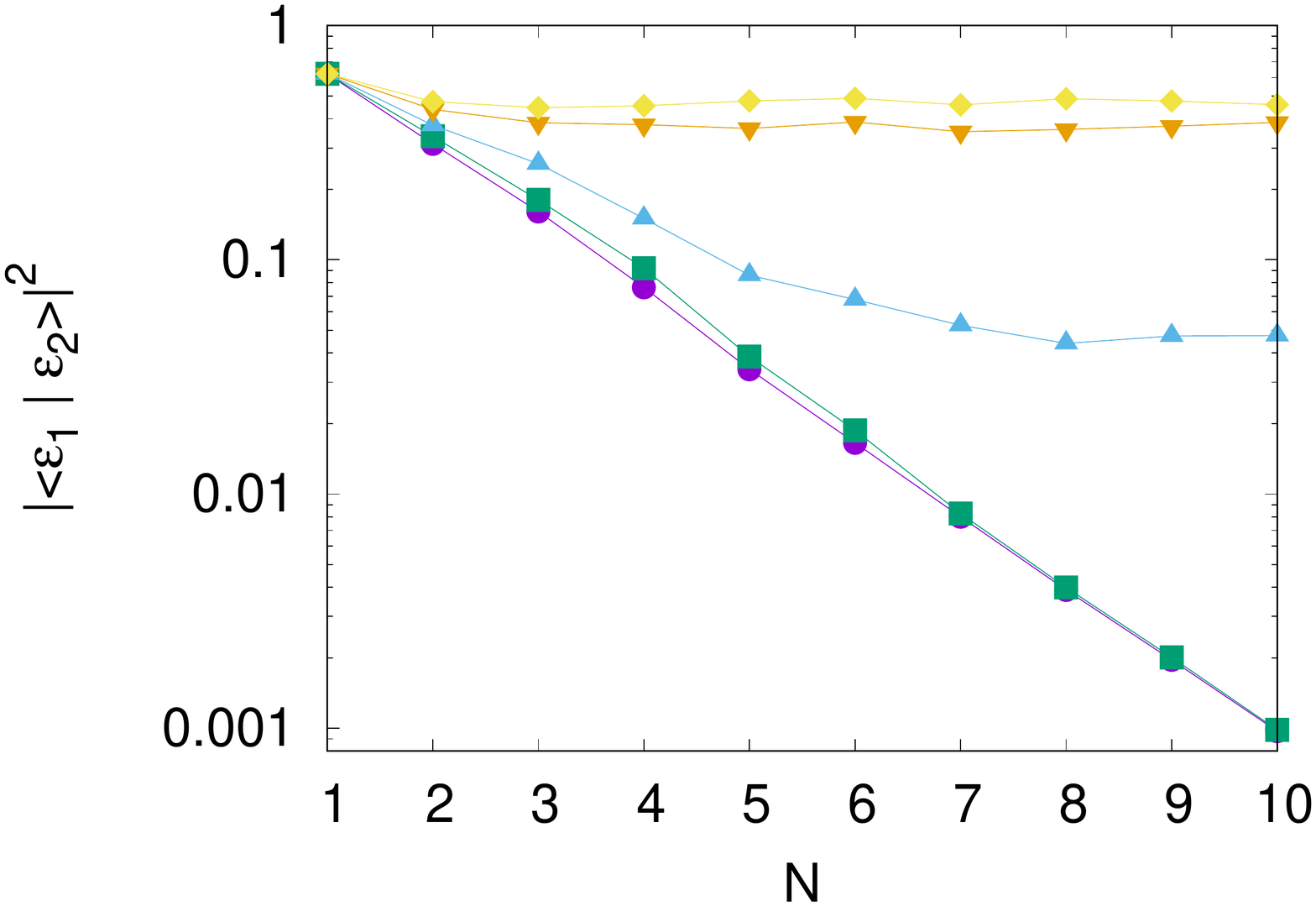} 
  \caption{Finite-size scaling for the long-time average of $\left|
    \braket{\varepsilon_1(t)}{\varepsilon_2(t)} \right|^2$, as a
    function of the number of qbits composing the environment,
    $N$. Solid circles represent the case $\alpha=0$; solid squares,
    $\alpha=0.5$; solid upper triangles, $\alpha=1$; solid lower
    triangles, $\alpha=2$, and solid diamons, $\alpha=4$.}
  \label{fig:caos}
\end{figure}

In Fig. \ref{fig:caos} we show how the long-time average of $\left|
\braket{\varepsilon_1(t)}{\varepsilon_2(t)} \right|^2$, calculated for
$2 \leq t \leq 50$, scales with the number of environmental qbits,
$N$, for five different values of $\alpha=0$, $0.5$, $1$, $2$, and
$4$. The results are averaged over $50$ different realizations. It is
clearly seen that the two fully chaotic cases, $\alpha=0$ (circles)
and $\alpha=0.5$ (sqares), behave in the same way; the overlap $\left|
\braket{\varepsilon_1(t)}{\varepsilon_2(t)} \right|^2$ decreases with
the number of environmental qbits, and therefore we can expect
$\ket{\varepsilon_1(t)}$ and $\ket{\varepsilon_2(t)}$ to become
ortogonal if the environment is large enough. The behavior of the case
with $\alpha=1$ (upper triangles) is different. First, the overlap
$\left| \braket{\varepsilon_1(t)}{\varepsilon_2(t)} \right|^2$
decreases with $N$, but it seems to reach an asymptotic value for $N
\gtrsim 7$. This fact suggests that a fully chaotic
apparatus-environment interaction is required for the scenario
described by the decoherence framework. This conclusion is reinforced
with the results for $\alpha=2$ (lower triangles) and $\alpha=4$
(diamonds). These two cases correspond with (almost) integrable
Hamiltonians, and their overlaps $\left|
\braket{\varepsilon_1(t)}{\varepsilon_2(t)} \right|^2$ remain large
independently of the number of environmental qbits.

\subsection{Summary of results}

The results discussed in the previous section narrow down the
circumstances under which Wigner's friend experiments are feasible, if
we take into account the decoherence interpretation of quantum
measurements. First, laboratories in which all the agents work must
have the structure given in Tab. \ref{tab:laboratorio}. Second, if
external agents want to perform interference experiments relying on
just two basis states, requirements listed in Tab. \ref{tab:protocol}
are mandatory. And third, if such circumstances hold, then the facts
F1 and F2 listed in Tab. \ref{tab:facts} characterize such
experiments. Fact F1 establishes that an observer cannot get a
conclusion about the exact state of the whole system (including
itself!) just from the outcome of as many experiments as it can
perform. On the contrary, the very fact of observing a definite
outcome entails that the observer is a part of a larger, entangled
superposition state, including an environment from which the observer
cannot get information. (Note that the decoherence framework establishes that this happens in any quantum measurement, independently of the existence of an external observer). Fact F2 refers to the practical consequences
of F1. It entails that all the agents involved in an experiment are
limited to discuss about the outcomes they obtain, outcomes that
depend both on their measuring apparatus {\em and the environmental
  degrees of freedom which have been traced out}. If either the
apparatus or the environmental degrees of freedom are different, then
the whole experiment is also different, and thus different outcomes
can be expected.

\begin{table}
  \begin{tabular}{p{0.5cm}p{7.5cm}}
    \hline
        {\bf F1} & After the measurement performed by agent $I$ is
    completed, the real state of the measured system, the measuring
    apparatus and the surrounding environment (which includes the
    agent itself) is given by Eq. (\ref{eq:environment1}), with
    $\left| \braket{\varepsilon_1(t)}{\varepsilon_2(t)} \right|^2 \sim
    0$.\\

{\bf F2} & All the results obtained by the agent $I$ are compatible with the mixed state given by Eq. (\ref{eq:colapso}). That is, it sees the system as if it were collapsed onto one of the possible outcomes of its experiment, despite fact F1.\\\hline
  \end{tabular}
  \caption{Summary of the facts consequence of the decoherence interpretation of quantum measurements, for Wigner's friend experiments.}
  \label{tab:facts}
  \end{table}

\section{Standard Wigner's friend experiments and the decoherence framework}
\label{sec:wigner}

In this section we discuss the consequences of the decoherence
framework in the standard Wigner's friend experiment
\cite{Wigner:61}. This discussion sets the grounds to analyze the extended versions of the experiments \cite{Renner:18,Bruckner:18,Proietti:19}.

Let us consider that an internal agent $I$ has performed a measurement
on an initial state given by Eq. (\ref{eq:decoherencia}). As we have
explained above, independently of the outcome it observes, the
resulting state is given by Eq. (\ref{eq:environment1}), which is the
result of the unitary evolution due to Hamiltonians \eqref{eq:medida}
and \eqref{eq:interaccion}. To simplify the notation, we consider the
whole state of the laboratory as follows,
\begin{subequations}
  \begin{align}
  \ket{h(t)} &= \ket{h}\ket{A_h}\ket{\varepsilon_1(t)}, \\
  \ket{v(t)} &= \ket{v}\ket{A_v}\ket{\varepsilon_2(t)},
  \end{align}
\end{subequations}
where both $\ket{h(t)}$ and $\ket{v(t)}$ may in general change with
time. Thus, the state after the measurement by agent $I$ is
\begin{equation}
  \ket{\Psi_1(t)} = \coef{1}{2} \left( \ket{h(t)} + \ket{v(t)} \right).
  \label{eq:wigner1}
\end{equation}

Following the protocol proposed by Wigner \cite{Wigner:61}, an
external agent, $E$, performs a measurement on $\ket{\Psi_1(\tau)}$,
at a particular instant of time $\tau$. Let us consider that the four
requisites, R1-R4, of Tab. \ref{tab:protocol} are fulfilled, and
therefore an interference experiment can be performed with a two-state
basis, $\left\{ \ket{\alpha(\tau)}, \ket{\beta(\tau)} \right\}$, given
by
\begin{subequations}
  \begin{align}
  \ket{\alpha(\tau)} &= \sin \theta \ket{h(\tau)} + \cos \theta \ket{v(\tau)}, \\
  \ket{\beta(\tau)} &= -\cos \theta \ket{h(\tau)} + \sin \theta \ket{v(\tau)},
  \end{align}
\end{subequations}
for an arbitrary value of the angle $\theta$. In this basis, the state $\ket{\Psi_1(\tau)}$ reads,
\begin{equation}
  \begin{split}
    \ket{\Psi_1(\tau)} &= \coef{1}{2}\left( \sin \theta + \cos \theta \right) \ket{\alpha(\tau)} + \\ &+ \coef{1}{2}\left( \sin \theta - \cos \theta \right) \ket{\beta(\tau)}.
    \end{split}
\end{equation}
Therefore, following the decoherence formalism, and as a consequence
of the same kind of unitary evolution than before, the state resulting
from agent $E$ measurement is
\begin{equation}
  \begin{split}
    \ket{\Psi_2(\tau)} &= \coef{1}{2}\left( \sin \theta + \cos \theta \right) \ket{\alpha(\tau)} \ket{A'_{\alpha}} \ket{\varepsilon'_1(\tau)} + \\ &+ \coef{1}{2}\left( \sin \theta - \cos \theta \right) \ket{\beta(\tau)} \ket{A'_{\beta}} \ket{\varepsilon'_2(\tau)},
    \end{split}
\end{equation}
where $A'$ represents its apparatus, and $\varepsilon'$ the
environment required by the decoherence framework.

Up to now, we have considered that both the measurement and the
correlation between the apparatus $A'$ and the environment
$\varepsilon'$ happen at time $\tau$. But this consideration is not
relevant. Taking into account that both the internal, $A$, and
the external, $A'$, apparati are continuously monitorized by their respective
environments, the former state unitarily evolves with a Hamiltonian
$H=H_I \otimes I_E + I_I \otimes H_E$, where $I_I$ ($I_E$) represents
the identity operator for the internal (external)
laboratory. Therefore, in any moment after the measurement the
resulting state is
\begin{equation}
  \begin{split}
    \ket{\Psi_2(t)} &= \coef{1}{2}\left( \sin \theta + \cos \theta \right) \ket{\alpha(t)} \ket{A'_{\alpha}} \ket{\varepsilon'_1(t)} + \\ &+ \coef{1}{2}\left( \sin \theta - \cos \theta \right) \ket{\beta(t)} \ket{A'_{\beta}} \ket{\varepsilon'_2(t)},
  \end{split}
  \label{eq:wigner_real}
\end{equation}
with $\left| \braket{\varepsilon'_1(t)}{\varepsilon'_2(t)} \right|^2
\sim 0$. And hence, any further experiment performed by agent $E$, in
which the external environment is not measured, is compatible with the
state:
\begin{equation}
  \begin{split}
    \rho_E &= \frac{1}{2} \left( \sin \theta + \cos \theta \right)^2 \ket{\alpha(t)} \ket{A'_{\alpha}} \bra{A'_{\alpha}} \bra{\alpha(t)} + \\
    & + \frac{1}{2} \left( \sin \theta - \cos \theta \right)^2 \ket{\beta(t)} \ket{A'_{\beta}} \bra{A'_{\beta}} \bra{\beta(t)}.
  \end{split}
  \label{eq:agentE}
\end{equation}

Two remarks are useful at this point. First, as we have pointed out
above, the real state of the system is given by
Eq. (\ref{eq:wigner_real}); the mixed state given by
Eq. (\ref{eq:agentE}) is only a description of what agent $E$ {\em
  sees}, that is, of what agent $E$ can infer from any further
measurements performed by itself, and what it is recorded in its
memory. Second, the interpretation of Eq. (\ref{eq:agentE}) is
independent of the precise forms of $\ket{\alpha(t)}$ and
$\ket{\beta(t)}$. The fact that the internal laboratory changes with
time, as a consequence of the monitorization by its environment, due
to the Hamiltonian \eqref{eq:interaccion}, has no influence on agent
$E$ conclusions because its apparatus remains pointing at either
$\alpha$ or $\beta$.

As we have explained in the previous section, the main difference
between the decoherence framework and the standard interpretation of
quantum measurements, consisting just in a correlation between the
system and the apparatus, is that definite outcomes arise as a
consequence of the environmental monitorization, given by the
Hamiltonian \eqref{eq:interaccion}, and therefore can be exactly
tracked at any instant of time. This fact releases us from the need of
choosing a particular perspective to intepret the results without
inconsistencies, as it is proposed in \cite{Baumann:19}; within the
decoherence framework, we just need to calculate the state of the
agents' memories. So, as the action of an external observer includes an
interaction with the internal environment, one may wonder the
consequences of such an action.  The measurement performed by agent
$E$ has changed the state of the system from
\begin{equation}
  \ket{\Psi_1(t)} = \coef{1}{2} \left( \ket{h(t)} + \ket{v(t)} \right)
  \otimes \ket{A'_0} \ket{\varepsilon'_0},
  \label{eq:wigner_inicial}
\end{equation}
where $\ket{A'_0}$ and $\ket{\varepsilon'_0}$ are the (irrelevant)
initial states of agent $E$ apparatus and the external environment, to
Eq. (\ref{eq:wigner_real}). The decoherence framework establishes that
agent $I$ sees the system as if it were collapsed either onto $\ket{h}$
or $\ket{v}$ (both with probability $p_h = p_v = 1/2$) as a
consequence of tracing out the degrees of freedom of $\varepsilon$,
$A'$ and $\varepsilon'$ from Eq. (\ref{eq:wigner_inicial}). But, as
the global state has changed onto Eq. (\ref{eq:wigner_real}) as a
consequence of agent $E$ measurement, a change of how agent $I$
perceives the reality is possible.  To answer this question, we can
rewrite Eq. (\ref{eq:wigner_real}) using the basis $\left\{ \ket{h},
\ket{v} \right\}$. The resulting state is
\begin{equation}
  \begin{split}
    \ket{\Psi_2(t)} &= \frac{\sin \theta}{\sqrt{2}} \left( \sin \theta + \cos \theta \right) \ket{h}\ket{A_h}\ket{\varepsilon_1(t)}\ket{A'_{\alpha}}\ket{\varepsilon'_{1}(t)} + \\
    &+ \frac{\cos \theta}{\sqrt{2}} \left( \sin \theta + \cos \theta \right) \ket{v}\ket{A_v}\ket{\varepsilon_2(t)}\ket{A'_{\alpha}}\ket{\varepsilon'_{1}(t)} + \\
    &+ \frac{\cos \theta}{\sqrt{2}} \left( \cos \theta - \sin \theta \right) \ket{h}\ket{A_h}\ket{\varepsilon_1(t)}\ket{A'_{\beta}}\ket{\varepsilon'_{2}(t)} + \\
    &+ \frac{\sin \theta}{\sqrt{2}} \left( \sin \theta - \cos \theta \right) \ket{v}\ket{A_v}\ket{\varepsilon_2(t)}\ket{A'_{\beta}}\ket{\varepsilon'_{2}(t)}.   
    \end{split}
  \end{equation}
As any further measurements performed by agent $I$ will involve
neither its environment, $\varepsilon$, nor agent $E$ apparatus, $A'$,
nor agent $E$ environment, $\varepsilon'$, the resulting outcomes can
be calculated tracing out all these three degrees of freedom. The
result is
\begin{equation}
  \begin{split}
    \rho_I &= \frac{1}{4} \left( 2 - \sin 4 \theta \right) \ket{h}\ket{A_h} \bra{h} \bra{A_h} + \\ &+ \frac{1}{4} \left( 2 + \sin 4 \theta \right) \ket{v}\ket{A_v} \bra{v} \bra{A_v}.
    \end{split}
  \label{eq:agentI}
\end{equation}

This is the first remarkable consequence of the decoherence framework,
and shows that one has to be very cautious when testing claims made at
different stages of an external interference experiment.
%
Let us imagine that the protocol
discussed in this section, with $\theta=\pi/8$, has been performed a
large number, $N$, of times. Then, let us suposse that we acceed to
the memory record of agent $I$ ---encoded in the pointer states of the
apparatus, $\ket{A_h}$ and $\ket{A_v}$--- {\em before} the external
interference measurement takes place, in every realization of the
experiment. This reading would reveal us that agent $I$ has observed $h$
roughly $N/2$ times, and $v$ roughly the same amount of times. Now,
let us imagine that an identical protocol is being performed by a
colleague, but with a slight difference: in every realization, she
reads the internal memory of agent $I$ {\em after} the external
interference measurement has been completed. Astonishingly, our
colleague's reading would reveal that agent $I$ has observed $h$
roughly $N/4$ times, and $v$ roughly $3N/4$ times. At a first sight,
this conclusion seems preposterous. Our colleague and we are reading
an identical internal memory of an identical quantum machine
performing an identical ensemble of experiments, modelled by identical
Hamiltonians, Eqs. \eqref{eq:medida} and \eqref{eq:interaccion}; but
we claim that the machine has observed $h$ $N/2$ times, and our
colleague claims that this outcome has occured only $N/4$ times. This
absurd contradiction is easily ruled out if we take into account that
the external measurement modifies the state of the internal
environment, which is the ultimate responsible of the definite
outcomes recorded on the memory of the internal agent, and therefore
it also modifies these records. Furthermore, the decoherence framework
provides an exact procedure to calculate these changes, as we have
pointed out above.

This significative result can be summarized by means of the following
statement: {\em if the internal agent $I$ observes a definite outcome,
  then the exernal interference measurement performed by agent $E$
  changes its memory record; if this change does not occur is because
  agent $I$ has not observed a definite outcome.}

The main conclusion we can gather from this analysis is that a
contradiction between two claims, one made before an external
interference measurement, and the other made afterwards, can be the
logical consequence of this interference measurement. Hence, the
arguments given in \cite{Renner:18}, which are based on the same kind
of contradictions, must be studied with care, taking into account all
the changes due to all the measurements performed throughout all the
protocol. This is the aim of the next section.

 To illustrate this analysis, we perform now a numerical simulation
covering all the protocol. We study the case with $\theta=\pi/8$, and
we consider that both environments are composed by $6$ qbits ---the
total size of the Hilbert space is $2^{15}=32768$. We start from the
state resulting from agent $I$ pre-measurement
\begin{equation}
  \ket{\Psi_0} = \coef{1}{2} \left( \ket{h} \ket{A_h} + \ket{v} \ket{A_v}  \right) \ket{\varepsilon_1} \ket{A'_{\alpha}} \ket{\varepsilon'_1},
 \label{eq:estado_inicial}
\end{equation}
where $\varepsilon_1$ and $\varepsilon'_1$ represent the first states
of the basis used to model the internal and the external environments,
respectively. Note that we have considered the state
$\ket{A'_{\alpha}}$ as the zero state of the apparatus, but the
results do not depend on this particular choice. From this state, the
system passes through three stages:


{\em Stage 1.-} From $t=0$ to $t=\tau_1$, the internal environment
interacts with apparatus $A$ to complete the measurement. Even though
the external agent $E$ has not performed any measurement yet, we also
consider a similar interaction for the external environment ---in such
a case, the external agent $E$ would see a definite outcome pointing
to zero, that in this case corresponds to the outcome $\alpha$. The
corresponding Hamiltonian is
\begin{equation}
\begin{split}  
  H_1 &= \left( \ket{A_h} \bra{A_h} \sum_{n,m} V^h_{nm} \ket{\varepsilon_n} \bra{\varepsilon_m} + \ket{A_v} \bra{A_v} \sum_{n,m} V^v_{nm} \ket{\varepsilon_n} \bra{\varepsilon_m} \right) \otimes I_E + \\
  &+ \left( \ket{A'_{\alpha}} \bra{A'_{\alpha}} \sum_{n,m} V^{\alpha}_{nm} \ket{\varepsilon'_n} \bra{\varepsilon'_m} + \ket{A'_{\beta}} \bra{A'_{\beta}} \sum_{n,m} V^{\beta}_{nm} \ket{\varepsilon'_n} \bra{\varepsilon'_m} \right) \otimes I_I,
  \end{split}
  \label{eq:stage1}
\end{equation}
where $I_I$ represents the identity operator over the laboratory in
which agent $I$ lives, and $I_E$ the identity operator over the
degrees of freedom corresponding to $A'$ and $\varepsilon'$.

{\em Stage 2.-} From $t=\tau_1$ to $t=\tau_2$, agent $E$ performs its
pre-measurement. We consider that the interaction with the external
environment is switched off, to model that this part of the
measurement is purely quantum \cite{nota5}. However, the interaction
between the internal apparatus and the internal environment still
exists, because the monitorization is always present after a
measurement is completed. The corresponding Hamiltonian is
\begin{equation}
\begin{split}  
  H_2 &= \left( \ket{A_h} \bra{A_h} \sum_{n,m} V^h_{nm} \ket{\varepsilon_n} \bra{\varepsilon_m} + \ket{A_v} \bra{A_v} \sum_{n,m} V^v_{nm} \ket{\varepsilon_n} \bra{\varepsilon_m} \right) \otimes I_E + \\
  &+ g \ket{\beta(\tau_1)} \bra{\beta(\tau_1)} \left[ \ket{A'_{\alpha}} \bra{A'_{\alpha}} + \ket{A'_{\beta}} \bra{A'_{\beta}} - \ket{A'_{\alpha}} \bra{A'_{\beta}} - \ket{A'_{\beta}} \bra{A'_{\alpha}} \right] \otimes I_I.
  \end{split}
  \label{eq:stage2}
\end{equation}
It is worth remarking that the requirements R1-R4 of
Tab. \ref{tab:protocol} have been explicitely taken into account. The
interaction leading to agent $E$ pre-measurement is based on
$\ket{\beta(\tau_1)}$, which is the exact state of the internal
laboratory at time $t=\tau_1$. The duration of this stage is exactly
$\tau_2 - \tau_1 = \pi/(2 g)$.

{\em Stage 3.-} From $t=\tau_2$ on, the external
environment gets correlated with apparatus $A'$, to complete the
measurement performed by agent $E$. Hence, the Hamiltonian is again
given by Eq. (\ref{eq:stage1}).


In summary, the system evolves from the initial state given by Eq. \eqref{eq:estado_inicial}, $\ket{\Psi_0}$, by means of $H_1$, given by
Eq. (\ref{eq:stage1}), from $t=0$ to $t=\tau_1$; by means of $H_2$,
given by Eq. (\ref{eq:stage2}), from $t=\tau_1$ to $t=\tau_2$; and by
means of $H_1$ again, from $t=\tau_2$ on. Agent's $I$ point of view is
directly obtained from the real state of the whole system,
$\ket{\Psi(t)}$, by tracing out the degrees of freedom corresponding to
$\varepsilon$, $A'$ and $\varepsilon'$. The resulting state can be written
\begin{equation}
  \begin{split}
    \rho_I (t) &= C_{hh} (t) \ket{h}\ket{A_h} \bra{h} \bra{A_h} + \\
    &+ C_{hv} (t) \ket{h}\ket{A_h} \bra{v} \bra{A_v} + \\
    &+ C_{vh} (t) \ket{v}\ket{A_v} \bra{h} \bra{A_h} + \\
    &+ C_{vv} (t) \ket{v}\ket{A_v} \bra{v} \bra{A_v}. 
  \end{split}
  \label{eq:internal}
\end{equation}
If $C_{hv} \sim 0$ and $C_{vh} \sim 0$, agent $I$ sees the system as
if it were collapsed onto either $\ket{h} \ket{A_h}$, with probability
$C_{hh}$, or $\ket{v} \ket{A_v}$, with probability $C_{vv}$ \cite{nota_colapso}.

Following the same line of reasoning, agent $E$ point of view is
obtained from $\ket{\Psi(t)}$ by tracing out the external environment,
$\varepsilon'$. The resulting state can be written
\begin{equation}
  \begin{split}
    \rho_E (t) &= C_{\alpha \alpha} (t) \ket{\alpha(t)}\ket{A_{\alpha}} \bra{\alpha(t)} \bra{A_{\alpha}} + \\
    &+ C_{\alpha \beta} (t) \ket{\alpha(t)}\ket{A_{\alpha}} \bra{\beta(t)} \bra{A_{\beta}} + \\
    &+ C_{\beta \alpha} (t) \ket{\beta(t)}\ket{A_{\beta}} \bra{\alpha(t)} \bra{A_{\alpha}} + \\
    &+ C_{\beta \beta} (t) \ket{\beta(t)}\ket{A_{\beta}} \bra{\beta(t)} \bra{A_{\beta}}. 
  \end{split}
  \label{eq:external}
\end{equation}
The interpretation is the same as before. If $C_{\alpha \beta} \sim
0$ and $C_{\beta \alpha} \sim 0$, agent $E$ sees the reality as it
if were collapsed onto either $\ket{\alpha(t)} \ket{A_{\alpha}}$, with
probability $C_{\alpha \alpha}$, or $\ket{\beta(t)} \ket{A_{\beta}}$,
with probability $C_{\beta \beta}$. It is worth to note that the
states of the internal laboratory $\ket{\alpha(t)}$ and
$\ket{\beta(t)}$, change with time, but this is not relevant for agent
$E$ point of view.

\begin{figure}
  \begin{tabular}{c}
    \includegraphics[width=0.5\columnwidth]{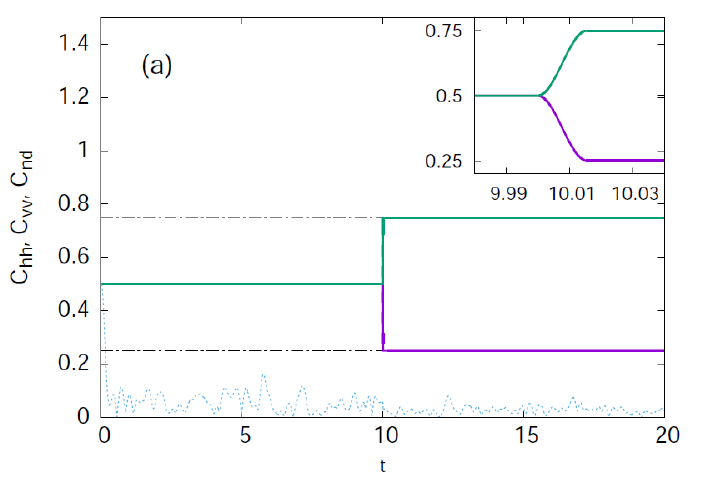} \\
    \includegraphics[width=0.5\columnwidth]{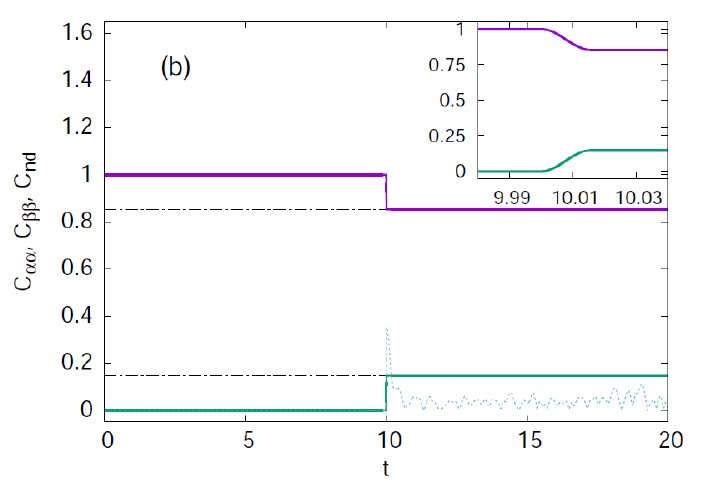}
    \end{tabular}
  \caption{Panel (a), matrix elements $C_{hh}$ (solid, violet line),
    $C_{vv}$ (solid green line), and $C_{nd}=\sqrt{\left|C_{hv}
      \right|^2 + \left|C_{vh} \right|^2}$ (dashed blue line), from
    Eq. (\ref{eq:internal}). Dotted-dashed lines show the expected
    values at stage 3. The inset show $C_{hh}$ and $C_{vv}$ around
    stage 2. Panel (b), matrix elements $C_{\alpha \alpha}$ (solid,
    violet line), $C_{\beta \beta}$ (solid green line), and
    $C_{nd}=\sqrt{\left|C_{\alpha \beta} \right|^2 + \left|C_{\beta
        \alpha} \right|^2}$ (dashed blue line), from
    Eq. (\ref{eq:external}). Dotted-dashed lines show the expected
    values at stage 3. The inset show $C_{\alpha \alpha}$ and
    $C_{\beta \beta}$ around stage 2. The number of qbits of both
    environment is $N=6$, $g=10^2$, $\tau_1=10$, and $\tau_2 -
    \tau_1 = \pi/200$.}
  \label{fig:internal_external}
\end{figure}

\begin{figure}
    \includegraphics[width=0.5\columnwidth]{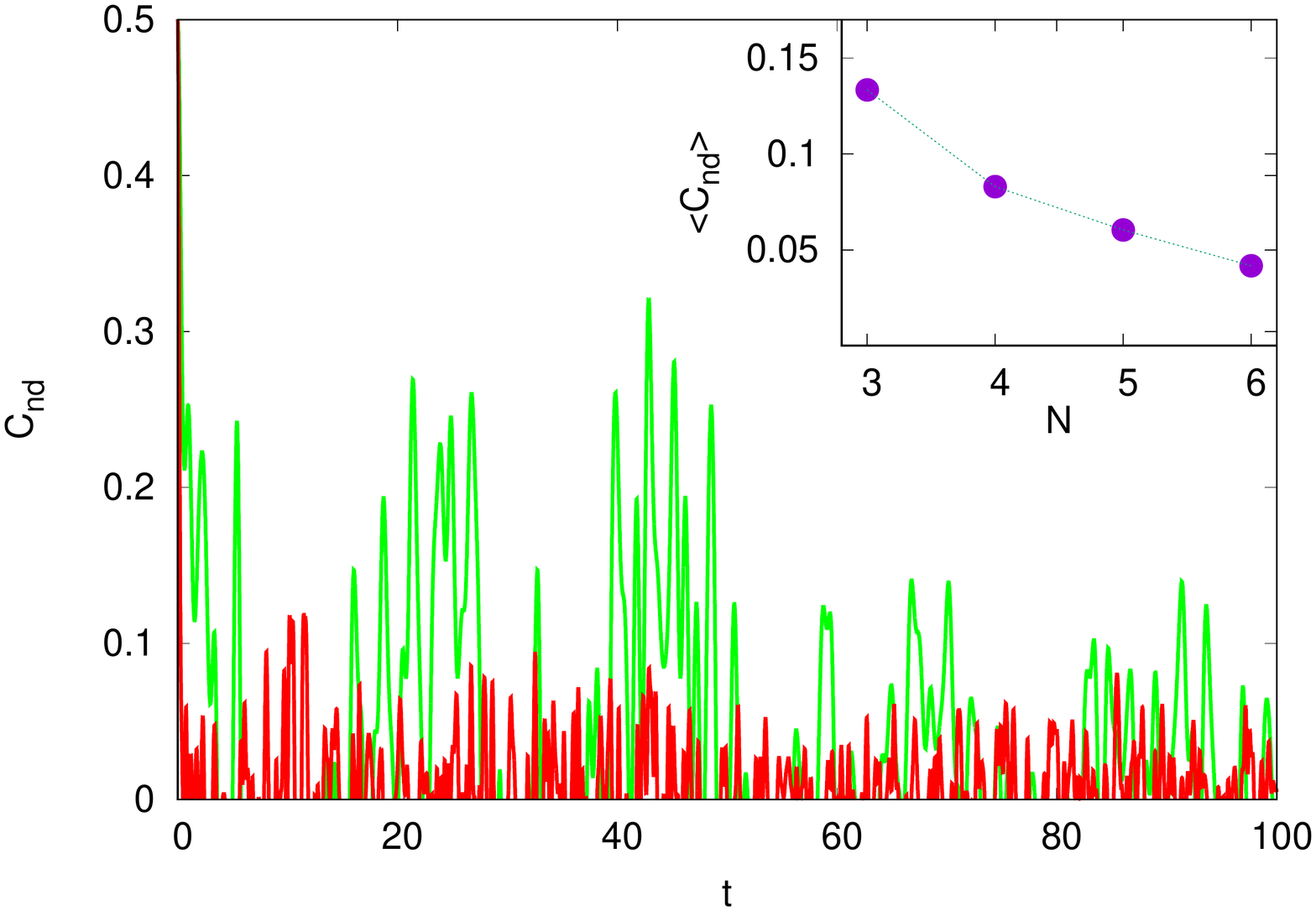}
  \caption{ $C_{nd}=\sqrt{\left|C_{hv} \right|^2 + \left|C_{vh}
      \right|^2}$, from Eq. (\ref{eq:internal}), for an environment with $N=3$ qbits (light green
    line) (dashed blue line), and for an environment with $N=6$ qbits
    (dark red line). In the inset, scaling analysis for the time
    average of $C_{nd}$ obtained from $t=3$ to $t=100$. }
  \label{fig:ruido}
\end{figure}

In panel (a) of Fig. \ref{fig:internal_external} we show the results
from agent's $I$ point of view. The coupling constant is set $g=100$;
$\tau_1=10$, and $\tau_2 - \tau_1 = \pi/200$. The non-diagonal
element, $C_{nd}=\sqrt{\left|C_{hv} \right|^2 + \left|C_{vh}
  \right|^2}$ (dotted blue line), is significatively large only at the
beginning of the simulation; from results in
Fig. \ref{fig:ortonormal}, we expect that larger environments give
rise to smaller values for $C_{nd}$ (see Fig. \ref{fig:ruido} for a
deeper discussion). Hence, our first conclusion is that agent's $I$
point of view is compatible with the photon collapsing either to
horizontal or to vertical polarizations. The measurement performed by
agent $E$, that starts at $\tau_1=10$, does not alter this
fact. However, as we clearly see in the inset of the same panel, this
measurement does change elements $C_{hh}$ (violet line) and $C_{vv}$
(green line). In the main part of the panel, we display the expected
values, given in Eq. (\ref{eq:agentI}), $C_{hh}=1/4$, $C_{vv}=3/4$, as
black dashed-dotted lines; we can see that these values are fast
reached. Furthermore, we can also see in the inset that this is a
smooth change, due to the physical interaction between the laboratory
and the apparatus $A'$. Therefore, {\em agent's $I$ point of view
  continuously changes during this small period of time}. As we have
already pointed out, the dependence of the Hamiltonian
\eqref{eq:stage2} on the internal environmental states alter the
definite outcomes observed by agent $I$, and therefore the records of
its internal memory. Thus, this simulation illustrates how the
apparent contradiction discussed above is solved.

Panel (b) of Fig. \ref{fig:internal_external} represents agent's $E$
point of view. Before performing the measurement, its apparatus
points $\alpha$ because this is chosen as zero. Then, at $t=\tau_1$
this point of view starts to change. $C_{\alpha \alpha}$ (solid violet
line) changes to $C_{\alpha \alpha} = 0.854$, the expected
value from Eq. (\ref{eq:agentE}), and equally $C_{\beta \beta}$ (solid
green line) changes to $C_{\beta \beta}=0.146$. During the first
instants of time after the pre-measurement, the non-diagonal element
$C_{nd}=\sqrt{\left|C_{\alpha \beta} \right|^2 + \left|C_{\beta
    \alpha} \right|^2}$ (blue dotted line) is significatively
different from zero; but, after the external environment has played
its role, agent $E$ point of view becomes compatible with the
laboratory collapsed either to $\alpha$ (with probability $p=0.854$)
or to $\beta$ (with probability $p=0.146$) as expected.

A finite-size scaling analysis of the non-diagonal element of $\rho_I$
is given in Fig. \ref{fig:ruido}. Due to the huge size of the whole
Hilbert space, it is not possible to reach large environmental
sizes. However, we clearly see in the inset how the size of this
non-diagonal element, $C_{nd}$, averaged from $t=3$ to $t=100$, decays
with the number of environmental qbits. Furthermore, a visual
comparison between the cases with $N=3$ (green line) and $N=6$ (red
line), given in the main panel of the same figure, corroborates this
impression. Therefore, we can conjecture that both agents $I$ and $E$
see their measured systems as if they were collapsed, provided that
their corresponding environments are large enough.

\begin{figure}
    \includegraphics[width=0.5\columnwidth]{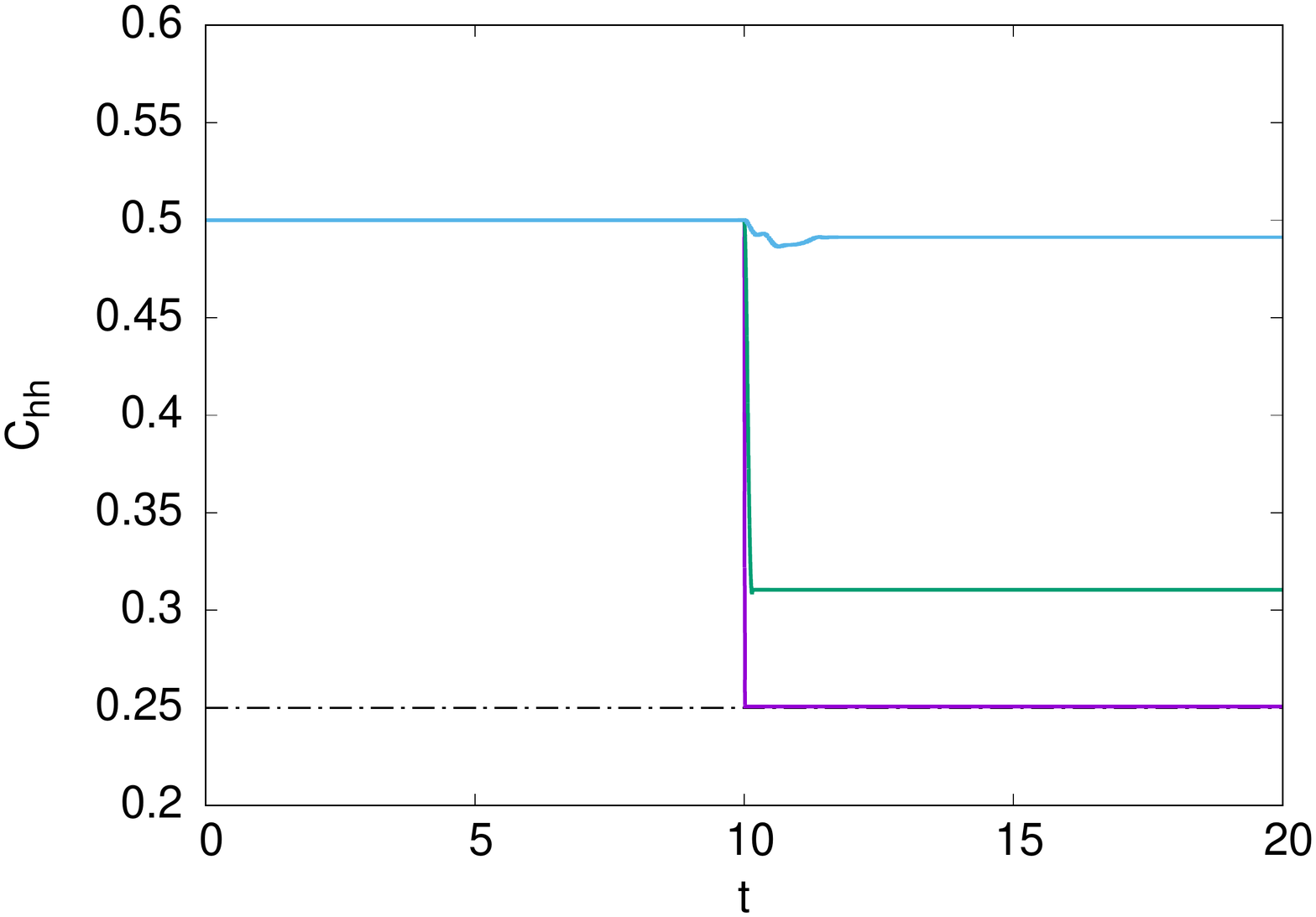}
  \caption{$C_{hh}$ from Eq. (\ref{eq:internal}) for different
    coupling constants $g$ in Eq. (\ref{eq:stage2}): $g=1$ (blue
    line), $g=10$ (green line), $g=100$ (violet line). Dotted-dashed
    line shows the expected value for stage 3.}
  \label{fig:internal2}
\end{figure}

Finally, we study how the results depend on the coupling constant
between the external apparatus, $A'$, and the laboratory whose state
is measured by agent $E$. In Fig. \ref{fig:internal2} we show $C_{hh}$
for $N=6$ and $g=1$ (blue line), $g=10$ (green line), and $g=100$
(violet line), together with the expected value, $C_{hh}=1/4$
(dotted-dashed black line). We conclude that this expected value is
reached only if $g$ is large enough. The explanation is quite
simple. If $g$ is small, the time required for the external apparatus
$A'$ to complete the pre-measurement is large compared with the
characteristic correlation time of the laboratory, given in
Fig. \ref{fig:correlacion}. Therefore, the state $\beta(\tau_1)$, used
in Eq. (\ref{eq:stage2}), ceases to be the real state of the
laboratory {\em while the external apparatus, $A'$, is still
  performing the pre-measurement}. As a consequence, the resulting
measurement is not correct, and neither agent $E$ nor agent $I$ reach
the expected results. This is an important fact that difficults a bit
more the external interference measurements trademark of Wigner's
friend experiments. Besides the requirements R1-R4 of
Tab. \ref{tab:protocol}, it is also mandatory that the external
pre-measurement is shorter than the characteristic time of the
internal dynamics of the measured laboratory. As it is shown in
Fig. \ref{fig:correlacion}, the larger the internal environment, the
shorter this time. Hence, if agent $I$ is a conscious (human) being,
composed by a huge number of molecules, the external interference
pre-measurement must be completed in a tiny amount of time.

The first conclusion we can gather from all these results is that,
according to the decoherence framework, {\em the memory records of all
  the agents involved in a Wigner's friend experiment will generically
  change after the actions of any other agents}, and therefore we must take these changes into account when comparing claims made at different stages of the experiment. As we will see in
next sections, this is the clue to interpret the extended versions of
the experiment.

Notwithstanding, agent $I$ still sees the reality as if the measured
photon were either horizontally or vertically polarised ---not in a
superposition of both states. Even more, states $\rho_E$, given by
Eq. (\ref{eq:agentE}), and $\rho_I$, given by Eq. (\ref{eq:agentI}),
seem incompatible at a first sight. But this is just a consequence of
the differences between the experiments performed by these two
agents. Agent $I$ sees the universe as if it were in state $\rho_I$,
because it ignores $\varepsilon$, $A'$ and $\varepsilon'$. On the
other hand, agent $E$ sees the universe as if it were in state
$\rho_E$, because it just ignores $\varepsilon'$, and therefore has
relevant information about $A'$ and $\varepsilon$. And, even more
important, both agents agree that their perceptions about the reality
are linked to the limitations of their experiments, and that the real
state of the universe is a complex, entangled and superposition state
involving the measured photon, both their apparatus, both the
environments that surround them, and themselves ---neither $\rho_I$,
nor $\rho_E$. Nevertheless, one of the most remarkable consequences of
the decoherence framework is that this fact does not prevent any of
the agents from making right claims about the outcomes observed by the
others. We will see in next sections that the four agents involved in
the extended version of the Wigner's friend experiment discussed in
Refs. \cite{Renner:18,Bruckner:18,Proietti:19} can make right ---not
contradictory--- claims about the other agents outcomes just
considering: {\em (i)} the results of their own experiments, that is,
the records of their own memories, and {\em (ii)} how external
interference measurements change these records. No other ingredients,
like the point of view change proposed in \cite{Baumann:19} are
required.

\section{Consistency of the quantum theory}
\label{sec:renner}

The aim of this section is to discuss the thought experiment proposed
in \cite{Renner:18} within the framework presented above. A number of
comments and criticisms have been already published, including
\cite{Bruckner:18} itself, and some others
\cite{Pusey:18,Salom:18,Healey:18,Lazarovici:19}. This work deals with
the original proposal in \cite{Renner:18}.

\subsection{No-go theorem and original interpretation}
\label{sec:scheme}

Both no-go theorems discussed in \cite{Renner:18,Bruckner:18} share a
similar scheme:

{\em (a)} A pair of entangled quantum systems is
generated. In \cite{Renner:18} it consists in a {\em quantum coin},
with an orthogonal basis given by $\{ \head$, $\tail \}$, and a
$1/2$-spin, spanned by $\{ \abajo$, $\arriba \}$. The initial
entangled state is
\begin{equation}
  \ket{\Psi} = \coef{1}{3} \head \abajo + \coef{2}{3} \tail \dcha,
  \label{eq:renner}
\end{equation}
where $\dcha = \coef{1}{2} \left( \abajo + \arriba \right)$.

To simplify the notation and make it compatible with
\cite{Bruckner:18,Proietti:19}, the following changes are made: {\em
  (i)} instead of the quantum coin and the spin in
Eq. (\ref{eq:renner}), two polarised photons are used; {\em (ii)} the
first photon is denoted by the subindex $a$, and the
second one, by the subindex $b$; {\em (iii)} the superpositions of
vertical and horizontal polarisation are denoted $\ket{+} =
\coef{1}{2} \left( \ket{h} + \ket{v} \right)$ and $\ket{-} = \coef{1}{2}
\left( \ket{h} - \ket{v} \right)$, respectively. With this notation, the initial
state in \cite{Renner:18} reads
\begin{equation}
  \ket{\Psi} = \coef{1}{3} \ket{h}_a \ket{v}_b + \coef{2}{3} \ket{v}_a \ket{+}_b.
  \label{eq:renner2}
\end{equation}

{\em (b)} Photon $a$ is sent to a closed laboratory $A$, and photon
$b$, to a closed laboratory $B$.

{\em (c)} An observer $I_A$, inside laboratory $A$, measures the state
of photon $a$; and an observer $I_B$, inside laboratory $B$, measures
the state of photon $b$.

{\em (d)} An external observer $E_A$ measures the state of the whole
laboratory $A$, and an external observer $E_B$ measures the state of
the whole laboratory $B$.

Both no-go theorems \cite{Renner:18,Bruckner:18} deal with the
observations made by $I_A$, $I_B$, $E_A$, and $E_B$. The one
formulated in \cite{Renner:18} is based upon the following
assumptions:

{\em Assumption Q.-} Let us consider that a quantum system is in the
state $\ket{\Psi}$. Then, let us suppose that an experiment has been
performed on a complete basis $\{ \ket{x_1}, \ldots, \ket{x_n} \}$,
giving an unknown outcome $x$. Then, if
$\expected{\Psi}{\pi_m}{\Psi}=1$, where $\pi_m = \ket{x_m}\bra{x_m}$,
for a particular state of the former basis, $\ket{x_m}$, then I am
certain that the outcome is $x=x_m$.

{\em Assumption C.-} If I am certain that some agent, upon reasoning
within the same theory I am using, knows that a particular outcome $x$ is
$x=x_m$, then I am also certain that $x=x_m$.

{\em Assumption S.-} If I am certain that a particular outcome is
$x=x_m$, I can safely reject that $x \neq x_m$.

The theorem says that there exist circumstances under which any quantum theory
satisfying these three assumptions is bound to yield constradictory
conclusions. The extended version of the Wigner's friend experiment
discussed in \cite{Renner:18} constitutes one paradigmatic example of
such circumstances.

Before continuing with the analysis, it is worth to remark that the
theorem focuses on particular outcomes that happen for certain ---with
probability $p=1$. It refers neither to the real state of the
corresponding system, nor to a subjective interpretation made by any
of the agents. Hence, its most remarkable feature is that
contradictions arise as consequences of simple observations.

Let us review now all the steps of the experiment from the four
agents' points of view. As it is explained in \cite{Renner:18}, to
infer their conclusions they need: {\em (i)} the knowledge of the
initial state of the whole system; {\em (ii)} their outcomes; {\em
  (iii)} the details of the experimental protocol, in order to predict
future outcomes, or track back past ones, relying on the unitary
evolutions of the corresponding (pre)measurements. We do not go into
details about the assumptions required to reach each conclusion; we
refer the reader to the original paper \cite{Renner:18} for that
purpose. Moreover, we do not consider now the decoherence framework;
all the measurements are understood as correlations between the
measured (part of the) system and the measuring apparatus.

{\em Step 1.-} Agent $I_A$ measures the initial state, given by
Eq. (\ref{eq:renner2}), in the basis $\left\{ \ket{h}_a, \ket{v}_a \right\}$.

{\em Fact 1:} Given the shape of the initial state, agent $I_A$
concludes that, if it obtains that photon $a$ is vertically polarised
(outcome $v_a$), then, a further measurement of the
laboratory $B$ in the basis $\{ \ket{+}_B, \ket{-}_B \}$ will lead
to the outcome $+_B$.

The resulting state of agent's $I_A$ measurement is
\begin{equation}
    \ket{\Psi}_1 = \coef{1}{3} \ket{h}_a \ket{v}_b \ket{A_h}_a + \coef{2}{3} \ket{v}_a \ket{+}_b \ket{A_v}_a.
  \label{eq:stage1b}
  \end{equation}

This expression can be simplified considering the whole state of the
laboratory $A$ which consists in the photon $a$ and the measuring
apparatus $A_a$. Hence, let
us denote
\begin{subequations}
  \begin{align}
  \ket{h}_A &\equiv \ket{h}_a \ket{A_h}_a, \\
  \ket{v}_A &\equiv \ket{v}_a \ket{A_v}_a.
  \end{align}
\end{subequations}
And, therefore, the state after this measurement is
\begin{equation}
  \ket{\Psi_1} = \coef{1}{3} \ket{h}_A \ket{v}_b + \coef{2}{3} \ket{v}_A \ket{+}_b.
  \label{eq:medida1}
  \end{equation}

Fact 1 {\em seems} compatible with this state. There is a perfect
correlation between state $\ket{v}_A$, which represents the case in
which agent $I_A$ has observed that the photon $a$ is vertically
polarised, and state $\ket{+}_b$. Thus, agent $I_A$ can deduce that
the laboratory $B$ will evolve from $\ket{+}_b$ to $\ket{+}_B$, as a
consequence of agent's $I_B$ measurement. And hence, considering
irrelevant the further action of agent $E_A$, because it does not deal
with laboratory $B$ \cite{Renner:18}, a further measurement on
laboratory $B$ will yield $+_B$, subjected to the outcome $v_a$. We
will see in Sec. \ref{sec:discussion} that considering irrelevant the
action of agent $E_A$ is not important if the decoherence framework is
{\em not} taken into account ---if the measurements consist just on
correlations between the systems and the apparati. In
Sec. \ref{sec:deco_renner} we will discuss how the decoherence
framework alter this fact.  

{\em Step 2.-} Agent $I_B$ measures photon $b$ in the basis $\{
\ket{h}_b, \ket{v}_b \}$.

{\em Fact 2:} If agent $I_B$ observes that the photon is horizontally
polarised, then the outcome of agent $I_A$ cannot correspond to a
horizontally polarised photon.

Using
the same notation as before (applied to laboratory $B$), the state
after agent $I_B$ completes its measurement is
\begin{equation}
    \ket{\Psi_2} = \coef{1}{3} \ket{v}_A \ket{h}_B + \coef{1}{3} \ket{v}_A \ket{v}_B + \coef{1}{3} \ket{h}_A \ket{v}_B.
  \label{eq:medida2}
\end{equation}

Therefore, there is a perfect correlation between $\ket{h}_B$ and
$\ket{v}_A$; the probability of observing $h_b$ and $h_a$ in the
same realization of the experiment is zero. Hence, all the previous
conclusions are well supported.

{\em Step 3.-} Agent $E_A$ measures laboratory $A$ in the basis $\{
\ket{+}_A, \ket{-}_A \}$, where
\begin{subequations}
  \begin{align}
  \label{eq:mas}
  \ket{+}_A &= \coef{1}{2} \left( \ket{h}_A + \ket{v}_A \right), \\
  \label{eq:menos}
  \ket{-}_A &= \coef{1}{2} \left( \ket{h}_A - \ket{v}_A \right).
  \end{align}
  \end{subequations}
Then, the state after this measurement is
\begin{equation}
\ket{\Psi_3} = \coef{2}{3} \ket{+}_A \ket{A'_+}_A  \ket{v}_B 
+ \coef{1}{6} \ket{+}_A \ket{A'_+}_A \ket{h}_B -
\coef{1}{6} \ket{-}_A \ket{A'_-}_A \ket{h}_B,
\label{eq:medida3}
\end{equation}
where $A'$ is the measuring apparatus used by agent $E_A$. From this state, we
obtain:

{\em Fact 3a:} If the outcome obtained by agent $E_A$ is $-_A$, then
agent $I_B$ has obtained an horizontally polarised photon, $h_b$, in
its measurement.

{\em Fact 3b}: Given facts 3a and 2, the outcome $-_A$, obtained by
agent $E_A$ determines that agent $I_A$ could not obtain an
horizontally polarised photon.

{\em Fact 3c}: Given the facts 3b and 1, the outcome $-_A$ determines
that a further measurement on laboratory $B$, in the basis $\{
\ket{+}_B, \ket{-}_B \}$ will necessary yield $+_B$.

The main conclusion we can infer from these sequential reasonings is
that, if agent $E_A$ observes $-_A$, then $E_B$ is bounded to observe
$+_B$. Therefore, it is not possible that outcomes $-_A$ and $-_B$
occur in the same realization of the experiment. Furthermore, as it is
discussed in detail in \cite{Renner:18}, relying on assumptions Q, S,
and C, it is straightforward to show that the four agents agree with
that.

The contradiction that (presumably) establishes that quantum theory
cannot consistently describe the use of itself consists in that the
probability of obtaining $-_A$ and $-_B$ in the same realization of
the experiments is $1/12$, {\em even though all the agents, relying on
  assumptions Q, C and S, agree that such probability must be zero}. This
can be easily inferred from the final state of the system after
measurements performed by all the agents (including $E_B$) are
completed,
\begin{equation}
  \begin{split}
    \ket{\Psi} &= \coef{3}{4} \ket{+}_A \ket{A'_+}_A \ket{+}_B \ket{A'_+}_B - 
\coef{1}{12} \ket{+}_A \ket{A'_+}_A \ket{-}_B \ket{A'_-}_B - \\ &- \coef{1}{12} \ket{-}_A \ket{A'_-}_A \ket{+}_B \ket{A'_+}_B - \coef{1}{12} \ket{-}_A \ket{A'_-}_A \ket{-}_B \ket{A'_-}_B.
  \end{split}
  \label{eq:cuatro}
\end{equation}

\subsection{The role of the decoherence framework}
\label{sec:deco_renner}

The first element that the decoherence framework introduces is that
every pre-measurement has to be fixed by the action of the
corresponding environment. Notwithstanding, this fact does not change
too much the equations discussed in the previous section. The states
of all the laboratories change with time, due to the continuous
monitorization by their environments, and all the measurements must be
completed at their exact times, following the results in
Tab. \ref{tab:protocol}, but the structure of all the resulting
equations is pretty much the same. For example, the state of
laboratory $A$ must be written
\begin{subequations}
  \begin{align}
    \label{eq:labA1}
    \ket{h(t)}_A &\equiv \ket{h}_a \ket{A_h}_a \ket{\varepsilon_1(t)}_a, \\
    \label{eq:labA2}
  \ket{v(t)}_A &\equiv \ket{v}_a \ket{A_v}_a \ket{\varepsilon_2(t)}_a,
  \end{align}
\end{subequations}
including the time-dependent environment, the final state of the whole
setup becomes
\begin{equation}
  \begin{split}
    \ket{\Psi} &= \coef{3}{4} \ket{+(\tau)}_A \ket{A'_+(\tau)}_A \ket{\varepsilon'_1(\tau)}_A \ket{+(\tau)}_B \ket{A'_+(\tau)}_B \ket{\varepsilon'_1(\tau)}_B - \\ &- 
\coef{1}{12} \ket{+(\tau)}_A \ket{A'_+(\tau)}_A \ket{\varepsilon'_1(\tau)}_A \ket{-(\tau)}_B \ket{A'_-(\tau)}_B \ket{\varepsilon'_2(\tau)}_B - \\ &- \coef{1}{12} \ket{-(\tau)}_A \ket{A'_-(\tau)}_A \ket{\varepsilon'_2(\tau)}_A \ket{+(\tau)}_B \ket{A'_+(\tau)}_B \ket{\varepsilon'_1(\tau)}_B - \\ &- \coef{1}{12} \ket{-(\tau)}_A \ket{A'_-(\tau)}_A \ket{\varepsilon'_2(\tau)}_A \ket{-(\tau)}_B \ket{A'_-(\tau)}_B \ket{\varepsilon'_2(\tau)}_B,
  \end{split}
  \label{eq:cuatro_deco}
\end{equation}
instead of much simpler Eq. (\ref{eq:cuatro}).

Another important point is that requisites R1-R4 from
Tab. \ref{tab:protocol}, together with the fast-enough realization of
the external interference experiments, are mandatory to reach the
previous conclusion. Hence, assumptions Q, S and C might only lead to
contradictory conclusions if the experiment is performed under very
specific circumstances. Results in Fig. \ref{fig:correlacion} suggest
that, the larger the laboratories $A$ and $B$ are, the more specific
the circumstances of the experiment must be. Thus, if agents are not
small quantum machines, composed by just a few qbits, but human
beings, composed by a huge number of particles, the probability that
such a contradiction might arise is virtually zero. Notwithstanding,
this conclusion only affects human beings acting as agents. Quantum
machines acting coherently, like the $53$ qbit quantum computer
recently developed \cite{google}, are free from this
limitation. Therefore, and despite the huge complexity of such
experiments, we can trust that they will be feasible in the future.

Now, let us imagine that quantum technologies are sufficently
developed, and let us go ahead with the experiment. That is, let us
wonder if quantum theory can consistently describe the use of itself,
relying on the decoherence formalism. For this purpose, we follow the
same guidelines of the setup in \cite{Renner:18}: we assume that all
agents are aware of the whole experimental procedure, and use the
decoherence framework to determine the outcomes that the other agents
have obtained or will obtain, conditioned to their own outcomes.

We start our analysis with fact 1. Considering the environmental
monitorization, the state after $I_A$ has performed its measurement is
\begin{equation}
  \ket{\Psi_1} = \coef{1}{3} \ket{h(t)}_A \ket{v}_b + \coef{2}{3} \ket{v(t)}_A \ket{+}_b,
  \label{eq:medida1_deco}
  \end{equation}
with $\ket{h(t)}_A$ and $\ket{v(t)}_A$ given by Eqs. (\ref{eq:labA1})
and (\ref{eq:labA2}). Let us study the conclusions that agent $I_A$ can reach from this state, relying on assumptions Q, C, and S, and the decoherence framework. After tracing out the environental degrees of freedom, Eq. (\ref{eq:medida1_deco}) gives rise to
\begin{equation}
  \rho_1 = \frac{1}{3} \ket{h}_a \ket{A_h}_a \ket{v}_b \bra{h}_a \bra{A_h}_a \bra{v}_b + \frac{2}{3} \ket{v}_a \ket{A_v}_a \ket{+}_b \bra{v}_a \bra{A_v}_a \bra{+}_b,
\end{equation}
that is, it establishes a correlation between the outcome $v_a$,
obtained by agent $I_A$, and the state $\ket{+}_b$. Hence, the first
conclusion that agent $I_A$ can reach is:

(i) If agent $I_B$ measures photon $b$ in the basis $\{ \ket{+}_b,
\ket{-}_b \}$, it will obtain the outcome $+_b$, if I have obtained
$v_a$.

However, as this measurement is not actually performed, this statement
is useless; agent $I_A$ needs further resonings and calculations to
reach a valid conclusion. Thus, it jumps to the next step in the
experimental protocol and takes into account the consequences of
agent's $I_B$ measurement. Relying on the decoherence framework and
considering the action of the corresponding unitary operators, agent
$I_A$ can calculate that the resulting state is
\begin{equation}
  \begin{split}
    \ket{\Psi_2} &= \coef{1}{3} \ket{v}_a \ket{A_v}_a \ket{\varepsilon_2(t)}_a  \ket{h}_b \ket{A_h}_b \ket{\varepsilon_1(t)}_b + \\ &+ \coef{1}{3} \ket{v}_a \ket{A_v}_a \ket{\varepsilon_2(t)}_a \ket{v}_b \ket{A_v}_b \ket{\varepsilon_2(t)}_b + \\ &+ \coef{1}{3} \ket{h}_a \ket{A_h}_a \ket{\varepsilon_1(t)}_a \ket{v}_b \ket{A_v}_b \ket{\varepsilon_2(t)}_b,
    \end{split}
\end{equation}
which can be written
\begin{equation}
    \ket{\Psi_2} = \coef{2}{3} \ket{v}_a \ket{A_v}_a \ket{\varepsilon_2(t)}_a \ket{+(t)}_B + \coef{1}{6} \ket{h}_a \ket{A_h}_a \ket{\varepsilon_1(t)}_a \ket{+(t)}_B - \coef{1}{6} \ket{h}_a \ket{A_h}_a \ket{\varepsilon_1(t)}_a \ket{-(t)}_B. 
\end{equation}
And, after tracing out the corresponding environmental degrees of freedom,
agent's $I_A$ perception can be written as
\begin{equation}
  \begin{split}
    \rho_2 &= \frac{2}{3} \ket{v}_a \ket{A_v}_a \ket{+(t)}_B \bra{v}_a \bra{A_v}_a \bra{+(t)}_B + \\ &+
    \frac{1}{6} \ket{h}_a \ket{A_h}_a \ket{+(t)}_B \bra{h}_a \bra{A_h}_a \bra{+(t)}_B + \frac{1}{6} \ket{h}_a \ket{A_h}_a \ket{-(t)}_B \bra{h}_a \bra{A_h}_a \bra{-(t)}_B - \\ &-
    \frac{1}{6} \ket{h}_a \ket{A_h}_a \ket{+(t)}_B \bra{h}_a \bra{A_h}_a \bra{-(t)}_B - \frac{1}{6} \ket{h}_a \ket{A_h}_a \ket{-(t)}_B \bra{h}_a \bra{A_h}_a \bra{+(t)}_B.
    \end{split}
  \end{equation}
Therefore, agent $I_A$ can make the following statement, which {\em
  seems similar} to fact 1:

(ii) If agent $I_B$ measures photon $b$ in the basis $\{ \ket{h}_b,
\ket{v}_b \}$, and subsequently, without any other measurement in
between, agent $E_B$ measures laboratory $B$ in the basis $\{
\ket{+(\tau)}_B, \ket{-(\tau)_B} \}$, the last one will obtain
$\ket{+}_B$, if I have obtained $\ket{v}_a$.

However, this statement does {\em not} represent the thought
experiment discussed in \cite{Renner:18}. The experimental protocol
establishes that agent $I_B$ measures photon $b$ in the basis $\{
\ket{h}_b, \ket{v}_b \}$, then agent $E_A$ measures the whole
laboratory $A$ in the basis $\{ \ket{+(\tau)}_A, \ket{-(\tau)_B} \}$,
and finally agent $E_B$ measures laboratory $B$ in the basis $\{
\ket{+(\tau)}_B, \ket{-(\tau)_B} \}$. This is the point at which the
differences between the decoherence framework and the standard
interpretations ---measurements as correlations between systems and
apparati--- emerge. As we have discussed in Sec. \ref{sec:wigner}, an
external interference measurement, like the one performed by agent
$E_A$, generally implies changes in the memory record of the measured
agent. Notwithstanding, as these changes can be exactly calculated,
agent $I_A$ can still rely on the decoherence framework to predict the
correlations between its outcome, $v_a$, and the one that agent $E_B$
will obtain when measuring laboratory $B$ in the basis $\{
\ket{+(\tau)}_B, \ket{-(\tau)}_B \}$. Just before the final
measurement by agent $E_B$, the state of the whole system can be
written
\begin{equation}
\begin{split}
  \ket{\Psi_3 (\tau)} &= \coef{3}{8} \ket{h}_a \ket{A_h}_a \ket{\varepsilon_1(\tau)}_a \ket{A'_+}_A \ket{\varepsilon_1'(\tau)}_A \ket{+}_B  +
  \coef{3}{8} \ket{v}_a \ket{A_v}_a \ket{\varepsilon_2(\tau)}_a \ket{A'_+}_A \ket{\varepsilon_1'(\tau)}_A \ket{+}_B - \\ &-
   \coef{1}{24} \ket{h}_a \ket{A_h}_a \ket{\varepsilon_1(\tau)}_a \ket{A'_+}_A \ket{\varepsilon_1'(\tau)}_A \ket{-}_B -
   \coef{1}{24} \ket{v}_a \ket{A_v}_a \ket{\varepsilon_2(\tau)}_a \ket{A'_+}_A \ket{\varepsilon_1'(\tau)}_A \ket{-}_B - \\ &-
   \coef{1}{24} \ket{h}_a \ket{A_h}_a \ket{\varepsilon_1(\tau)}_a \ket{A'_-}_A \ket{\varepsilon_2'(\tau)}_A \ket{+}_B +
   \coef{1}{24} \ket{v}_a \ket{A_v}_a \ket{\varepsilon_2(\tau)}_a \ket{A'_-}_A \ket{\varepsilon_2'(\tau)}_A \ket{+}_B - \\ &-
   \coef{1}{24} \ket{h}_a \ket{A_h}_a \ket{\varepsilon_1(\tau)}_a \ket{A'_-}_A \ket{\varepsilon_2'(\tau)}_A \ket{-}_B +
   \coef{1}{24} \ket{v}_a \ket{A_v}_a \ket{\varepsilon_2(\tau)}_a \ket{A'_-}_A \ket{\varepsilon_2'(\tau)}_A \ket{-}_B.
\end{split}
\label{eq:new_fact1}
\end{equation}
Hence, tracing out agent's $I_A$ environment, and both agent's $E_A$
apparatus and environment, since agent's $E_A$ outcome is irrelevant, the state of agent's $I_A$ memory reads
\begin{equation}
\begin{split}
  \rho_3 &= \frac{5}{12} \ket{h}_a \ket{A_h}_a \ket{+}_B \bra{h}_a \bra{A_h}_a \bra{+}_B +
  \frac{5}{12} \ket{v}_a \ket{A_v}_a \ket{+}_B \bra{v}_a \bra{A_v}_a \bra{+}_B + \\ &+
  \frac{1}{12} \ket{h}_a \ket{A_h}_a \ket{-}_B \bra{h}_a \bra{A_h}_a \bra{-}_B +
  \frac{1}{12} \ket{v}_a \ket{A_v}_a \ket{-}_B \bra{v}_a \bra{A_v}_a \bra{-}_B. 
\end{split}
\label{eq:new_fact1b}
\end{equation}
And therefore, relying on assumption Q, agent $I_A$ can make the
following claims: {\em (i)} the system is in state given by
Eq. (\ref{eq:new_fact1}) just before agent's $E_B$ measurement; {\em
  (ii)} a certain outcome in the basis $\{ \ket{+(\tau)}_B,
\ket{(-)}_B \}$ is going to be obtained; {\em (iii)} neither
$\expected{\Psi_3(\tau)}{\pi_{+,B}}{\Psi_3(\tau)}=1$, nor
$\expected{\Psi_3(\tau)}{\pi_{-,B}}{\Psi_3(\tau)}=1$, conditioned to
my memory record is $v_a$. Hence, the decoherence framework modifies fact
$1$, giving rise to

{\em New fact 1:} If agent $I_A$ obtains that the photon is vertically
polarised (outcome $v_a$), then, a further measurement of the
laboratory $B$ in the basis $\{ \ket{+(\tau)}_B, \ket{-(\tau)}_B \}$
will lead to either $+_B$ (with $p=5/6$) or $-_B$ (with $p=1/6$).

It is worth noting that this prediction can be experimentally
confirmed by simultaneously reading the memory records of agents $I_A$
and $E_B$, as soon as the last outcome is fixed by the corresponding
environmental monitorization. Even though it is reasonable to wonder
if this inconclusive statement is a consequence of the changes induced
in agent's $I_A$ memory by the external interference measurement
performed by agent $E_A$, the key point is that no correlations
between $I_A$ and $E_B$ perceptions exist before the action of agent
$E_A$, so there is no other way to determine whether agent's $E_B$
outcome is bounded by agent's $I_A$ or not. In
Sec. \ref{sec:discussion} we will see that the results are different
is the decoherence framework is not taken into account.

As the inconsistency discussed in \cite{Renner:18} is based on fact 1,
the result we have obtained is enough to show that the
decoherence framework is free from it. Notwithstanding, to delve in
the interpretation of this remarkable thought experiment, follows a
discussion about facts 2 and 3. 

Again, we focus on the predictions that the involved agents can make
by means of the decoherence framework, and their experimental
verification by reading their corresponding memory records. Fact 2 is made from agent's $I_B$ point of view, so we focus on the state of the system after
both agents $I_A$ and $I_B$ have completed their
measurements, which reads
\begin{equation}
  \begin{split}
    \ket{\Psi_2} &= \coef{1}{3} \ket{v}_a \ket{A_v}_a \ket{\varepsilon_2(t)}_a  \ket{h}_b \ket{A_h}_b \ket{\varepsilon_1(t)}_b + \\ &+ \coef{1}{3} \ket{v}_a \ket{A_v}_a \ket{\varepsilon_2(t)}_a \ket{v}_b \ket{A_v}_b \ket{\varepsilon_2(t)}_b + \\ &+ \coef{1}{3} \ket{h}_a \ket{A_h}_a \ket{\varepsilon_1(t)}_a \ket{v}_b \ket{A_v}_b \ket{\varepsilon_2(t)}_b.
    \end{split}
\end{equation}

To obtain their common view of the system, both their environment must be traced out. Hence, the memory records of both agents are compatible with the mixed state given by \cite{nota6}
\begin{equation}
\begin{split}
  \rho_2 &= \frac{1}{3} \ket{v}_a \ket{A_v}_a \ket{h}_b \ket{A_h}_b \bra{v}_a \bra{A_v}_a \bra{h}_b \bra{A_h}_b + \\
  &+ \frac{1}{3} \ket{v}_a \ket{A_v}_a \ket{v}_b \ket{A_v}_b \bra{v}_a \bra{A_v}_a \bra{v}_b \bra{A_v}_b + \\
  &+ \frac{1}{3} \ket{h}_a \ket{A_h}_a \ket{v}_b \ket{A_v}_b \bra{h}_a \bra{A_h}_a \bra{v}_b \bra{A_v}_b.
\end{split}
\label{eq:fact2}
  \end{equation}
That is, if agent $I_B$ relies on the decoherence framework to
calculate agent's $I_A$ outputs conditioned to the one it has
obtained, it can safely conclude fact 2 {\em at this stage of the
  experiment}.

Let us now proceed with fact 3. As it is formulated from agent's $E_A$
point of view, we start from the state of the system after agent's
$E_A$ measurement, which reads
\begin{equation}
\begin{split}
  \ket{\Psi_3 (\tau)} &= \coef{2}{3} \ket{+(\tau)}_A\ket{A'_+}_A \ket{\varepsilon'_1 (\tau)}_A \ket{v}_b \ket{A_v}_b \ket{\varepsilon_2(\tau)}_b + \\
  &+ \coef{1}{6} \ket{+(\tau)}_A\ket{A'_+}_A \ket{\varepsilon'_1 (\tau)}_A \ket{h}_b \ket{A_h}_b \ket{\varepsilon_1(\tau)}_b - \\
  &- \coef{1}{6} \ket{-(\tau)}_A\ket{A'_-}_A \ket{\varepsilon'_2 (\tau)}_A \ket{h}_b \ket{A_h}_b \ket{\varepsilon_1(\tau)}_b.
\end{split}
\label{eq:step3}
\end{equation}

Eq. (\ref{eq:step3}) represents the state of the whole system after
the measurements performed by agents $I_A$, $I_B$ and $E_A$ are
completed. The way that these agents perceive this state depends again
on the action of their respective environments, and therefore can be
described by tracing out the corresponding degrees of freedom. A joint vision of agents $I_B$ and $E_A$ is
obtained tracing out the environments $\varepsilon_b$ and
$\varepsilon'_A$, leading to
\begin{equation}
  \begin{split}
    \rho_3 (\tau) &= \frac{2}{3} \ket{+(\tau)}_A\ket{A'_+}_A \ket{v}_b \ket{A_v}_b \bra{+(\tau)}_A\bra{A'_+}_A \bra{v}_b \bra{A_v}_b + \\
    &+ \frac{1}{6} \ket{+(\tau)}_A\ket{A'_+}_A \ket{h}_b \ket{A_h}_b \bra{+(\tau)}_A\bra{A'_+}_A \bra{h}_b \bra{A_h}_b + \\
    &+ \frac{1}{6}\ket{-(\tau)}_A\ket{A'_-}_A \ket{h}_b \ket{A_h}_b \bra{-(\tau)}_A\bra{A'_-}_A \bra{h}_b \bra{A_h}_b.
  \end{split}
  \label{eq:fact3a}
\end{equation}
This state is fully compatible with fact 3a. At this stage of the
experiment, the correlation between the memory records of $I_B$ and
$E_A$ is incompatible with the outcomes $-_A$ and $v_b$ being obtained
at the same run of the experiment. This means that agent $E_A$ can use
assumption Q to conclude: {\em (i)} system is in state given by
Eq. (\ref{eq:step3}) after my measurement; {\em (ii)} a certain
outcome was obtained by agent $I_B$ in the basis $\{ \ket{h}_b,
\ket{v}_b \}$; {\em (iii)} as
$\expected{\Psi_3(\tau)}{\pi_{h,b}}{\Psi_3(\tau)}=1$, conditioned I
have obtained $-_A$, then fact 3a is correct.  Furthermore, agent
$I_B$, relying only on its outcome, the details of the whole protocol
and the decoherence framework, can also predict fact 3a.

To follow with the argument, agent $E_A$ performs a nested reasoning
to determine the outcome obtained by agent $I_A$. A very relevant
point is the time at which agent's $I_A$ memory record is
evaluated. If we re-write the current state of the system,
Eq. (\ref{eq:step3}), in a basis including $\left\{ \ket{h}_a
\ket{h}_b, \ket{h}_a \ket{v}_b, \ket{v}_a \ket{h}_b, \ket{v}_a
\ket{v}_b \right\}$, we obtain
\begin{equation}
  \begin{split}
    \ket{\Psi_3(\tau)} &= \coef{1}{3} \ket{h}_a \ket{A_h}_a \ket{\varepsilon_1(\tau)}_a \ket{A'_+}_A \ket{\varepsilon'_1(\tau)}_A \ket{v}_b \ket{A_v}_b \ket{\varepsilon_2(\tau)}_b + \\
    &+ \coef{1}{3} \ket{v}_a \ket{A_v}_a \ket{\varepsilon_2(\tau)}_a \ket{A'_+}_A \ket{\varepsilon'_1(\tau)}_A \ket{v}_b \ket{A_v}_b \ket{\varepsilon_2(\tau)}_b + \\
    &+  \coef{1}{12} \ket{h}_a \ket{A_h}_a \ket{\varepsilon_1(\tau)}_a \ket{A'_+}_A \ket{\varepsilon'_1(\tau)}_A \ket{h}_b \ket{A_h}_b \ket{\varepsilon_1(\tau)}_b + \\
    &+  \coef{1}{12} \ket{v}_a \ket{A_v}_a \ket{\varepsilon_2(\tau)}_a \ket{A'_+}_A \ket{\varepsilon'_1(\tau)}_A \ket{h}_b \ket{A_h}_b \ket{\varepsilon_1(\tau)}_b - \\
    &-  \coef{1}{12} \ket{h}_a \ket{A_h}_a \ket{\varepsilon_1(\tau)}_a \ket{A'_-}_A \ket{\varepsilon'_2(\tau)}_A \ket{h}_b \ket{A_h}_b \ket{\varepsilon_1(\tau)}_b + \\
    &+  \coef{1}{12} \ket{v}_a \ket{A_v}_a \ket{\varepsilon_2(\tau)}_a \ket{A'_-}_A \ket{\varepsilon'_2(\tau)}_A \ket{h}_b \ket{A_h}_b \ket{\varepsilon_1(\tau)}_b.
  \end{split}
  \label{eq:real3c}
\end{equation}
Thus, to determine the joint vision of agents $I_A$ and $I_B$ {\em at this
  stage of the experiment} we have just to trace out $\varepsilon_a$,
$\varepsilon_b$, $A'_A$ and $\varepsilon'_A$ from the density matrix
arising from this wavefunction. This leads to
\begin{equation}
  \begin{split}
    \rho_3 &= \frac{1}{3} \ket{h}_b \ket{A_h}_b \ket{v}_a \ket{A_v}_a \bra{h}_b \bra{A_h}_b \bra{v}_a \bra{A_v}_a + \\
    &+ \frac{1}{3} \ket{v}_b \ket{A_v}_b \ket{v}_a \ket{A_v}_a \bra{v}_b \bra{A_v}_b \bra{v}_a \bra{A_v}_a + \\
    &+ \frac{1}{6} \ket{h}_b \ket{A_h}_b \ket{h}_a \ket{A_h}_a \bra{h}_b \bra{A_h}_b \bra{h}_a \bra{A_h}_a + \\
    &+ \frac{1}{6} \ket{v}_b \ket{A_v}_b \ket{h}_a \ket{A_h}_a \bra{v}_b \bra{A_v}_b \bra{h}_a \bra{A_h}_a.
  \end{split}
  \label{eq:fact3c}
\end{equation}

This is one the most remarkable consequences of the decoherence
framework. In Sec. \ref{sec:wigner} we have shown that external
interference measurements generally change the memory records of
measured agents. Eq. (\ref{eq:fact3c}) shows that such interference
measurements also change the correlations between the memories of two
distant agents. If the correlations between agents's $I_A$ and $I_B$
outcomes are evaluated {\em before} the interference experiment
performed by agent $E_A$, fact 2 is correct; if they are evaluated
{\em afterwards}, it changes to: if agent $I_B$ has observed $h_b$,
then agent's $I_A$ memory record is compatible with both $h_a$ and
$v_a$. It is worth noting that the decoherence framework can be used
by all the agents to calculate {\em both} situations.

The agents involved in the thought experiment devised in
\cite{Renner:18} use the time evolution corresponding to each
measurement to track the system {\em back}, that is, in the language
of the decoherence framework, to calculate what agents $I_B$ and $I_A$
{\em thought in the past}. Hence, agent $E_A$ can rely on the
decoherence framework to conclude: {\em (i)} given
Eq. (\ref{eq:step3}), agent $I_B$ obtained the outcome $h_b$ {\em
  before} my own measurement, since no changes in laboratory $B$ have
ocurred in between; {\em (ii)} hence, independently of what agent
$I_A$ thinks now, it obtained the outcome $v_a$ before my own
measurement and conditioned to agent's $I_B$ outcome $h_b$; {\em
  (iii)} therefore, agent's $I_A$ memory record {\em was} $v_a$ in the
past, if I have obtained $-_A$, even though it can be either $v_a$ and
$h_a$ now.

The previous paragraph illustrates one of the most significatives
features of the decoherence framework: it can be used to calculate
both the past and the current state of all agents's memory records; no
ambiguities arise as a consequence of external interference
experiments. 

Regarding the thought experiment devised in \cite{Renner:18}, a proper
use of the decoherence framework, taking into account the exact times
at which the agents make their claims, shows that both facts 3a and 3b
are correct. But this framework also shows that fact 3c is {\em not}
correct, beacuse it relies on fact 1, which is incompatible with it.
Hence, the reasonings discussed in this section invalidate the proof
of the no-go theorem presented in \cite{Renner:18}. If assumptions Q,
S and C are used within the decoherence framework, agents $I_A$,
$I_B$, $E_A$ and $E_B$ do not reach the contradictory conclusion that
$-_A$ implies $+_B$. The key point in this argument is that one agent
must predict a correlation which is only fixed after an external
interference experiment on itself, if it wants to make a claim about
the final outcome of the protocol. The standard interpretation of
quantum measurements is ambiguous about this point. One can suspect
that something weird might happen, but a calculation to confirm or to
refute this thought cannot be done. On the contrary, the decoherence
framework provides exact results that can be tested by means of a
proper experiment

Finally, it is worth to remark that we have {\em not} proved that the
decoherence framework is free from inconsistencies. We have just shown
that the proof of the theorem proposed in \cite{Renner:18} is not
valid if the decoherence framework is taken into account. But the main
statement of the theorem can be still considered as a conjecture.

\subsection{Discussion}
\label{sec:discussion}

The conclusions of the previous section are enterely based on the
decoherence framework. Resuls of \cite{Renner:18} are well
substantiated if this framework is not taken into account, that is, if
a correlation between a system and a measuring apparatus is considered
enough to complete a measurement. In such a case, the final state of
the protocol can be written in four different shapes
\begin{subequations}
  \begin{align}
    \label{eq:alpha}
    \ket{\Psi_{\alpha}} &= \coef{1}{3} \ket{v}_A \ket{h}_B + \coef{1}{3} \ket{h}_A \ket{v}_B + \coef{1}{3} \ket{v}_A \ket{v}_B, \\
    \label{eq:beta}
    \ket{\Psi_{\beta}} &= \coef{2}{3} \ket{+}_A \ket{h}_B - \coef{1}{6} \ket{-}_A \ket{h}_B + \coef{1}{6} \ket{+}_A \ket{h}_B, \\
    \label{eq:gamma}
    \ket{\Psi_{\gamma}} &= \coef{2}{3} \ket{v}_A \ket{+}_B - \coef{1}{6} \ket{h}_A \ket{-}_B + \coef{1}{6} \ket{h}_A \ket{-}_B, \\
    \label{eq:delta}
    \ket{\Psi_{\delta}} &= \coef{3}{4} \ket{+}_A \ket{+}_B - \coef{1}{12} \ket{+}_A \ket{-}_B + \coef{1}{12} \ket{-}_A \ket{-}_B - \coef{1}{12} \ket{-}_A \ket{+}_B,    
    \end{align}
  \end{subequations}
relying on four different basis. If the preferred basis for each
measurement is {\em not} fixed by a unique mechanism, like the one
coming from the decoherence framework, the conclusions of the involved
agents become ambiguous. The following reasoning can be understood as a
consequence of the basis ambiguity problem \cite{Zurek:03}:

Eq. (\ref{eq:gamma}) can be used to establish a perfect correlation
between the outcomes $v_a$ and $+_B$: if laboratory $A$ is in state
$\ket{v}_A$, which can be understood as the state resulting from the
outcome $v_a$ obtained by agent $I_A$, then the outcome $+_B$ is
guaranteed. Hence, fact 1 is well supported ---the final state of the
whole experiment can be written in a way compatible with it. In a
similar way. Eq. (\ref{eq:beta}) establishes a perfect correlation
between $\ket{-}_A$ and $\ket{h}_B$, which can be interpreted as
follows: if agent $E_A$ as obtained $-_A$, then agent $E_B$ has
obtained $h_B$. Again, the final state of the whole protocol is
compatible with this fact. Finally. Eq. (\ref{eq:alpha}) establishes a
perfect correlation between $\ket{h_B}$ and $\ket{v_A}$, meaning that
if agent $I_B$ has obtained $h_B$, then agent $I_A$ has obtained
$v_A$. And this is again compatible with the final state of the
experiment.

Hence, as a consequence of the basis ambiguity problem, agents $I_A$,
$I_B$, $E_A$ and $E_B$ can rely on Eqs. (\ref{eq:alpha}),
(\ref{eq:beta}) and (\ref{eq:gamma}) to infer a conclusion
incompatible with Eq. (\ref{eq:delta}). As we have discussed in
Sec. \ref{sec:deco_renner}, the decoherence framework fixes this bug
by ruling out the basis ambiguity, and by providing just one preferred
basis for each outcome.

\section{Observer-independent facts}
\label{sec:bruckner}

This section deals with the no-go theorem discussed in
\cite{Bruckner:18}. This theorem has been experimentally confirmed in
\cite{Proietti:19}. A criticism is published in \cite{Healey:18}.

\subsection{Original version of the experiment and no-go theorem}

The structure of this experiment has been already discussed in
Sec. \ref{sec:scheme}. The only difference is the initial state, which
consists in a pair of polarised photons, spanned by $\{ \ket{h},
\ket{v} \}$, and reads
\begin{equation}
  \begin{split}
    \ket{\Psi}_{\beta} &= \coef{1}{2} \cos \frac{\pi}{8} \left( \ket{h}_a \ket{v}_b + \ket{v}_a \ket{h}_b \right) +  \\ &+ \coef{1}{2} \sin \frac{\pi}{8} \left(\ket{h}_a \ket{h}_b - \ket{v}_a \ket{v}_b \right).
  \end{split}
  \label{eq:bruckner2}
\end{equation}

This state is used to illustrate a no-go theorem that establishes that
the following four statements are incompatible, that is, are bounded to
yield a contradiction:

{\em Statement 1.-} Quantum theory is valid at any scale.

{\em Statement 2.-} The choice of the measurement settings of one
observer has no influence on the outcomes of other distant
observer(s).

{\em Statement 3.-} The choice of measurement settings is
statistically independent from the rest of the experiment.

{\em Statement 4.-} One can jointly assign truth values to the
propositions about outcomes of different observers.

In \cite{Bruckner:18, Proietti:19}, the thought experiment used to
proof this theorem consists of the following steps:

{\em (i)} The internal agents, $I_A$ and $I_B$, perform their
(pre)measurements, that is, establish a correlation between the
measured photons and their apparati given by
Eq. \eqref{eq:measurement2}.

{\em (ii)} The external agents, $E_A$ and $E_B$, choose between
performing interference experiments, or measuring the polarisation of
the internal photons.

{\em (iii)} A Bell-like test is performed on the four different
combinations resulting from point (ii), to conclude that it is not
possible to jointly assign truth values to the outcomes obtained by
the internal and the external agents.

This protocol was experimentally performed in \cite{Proietti:19},
validating the violation of the Bell-like test prediction in
\cite{Bruckner:18}.

The state resulting from step (i) is
\begin{equation}
\begin{split}
  \ket{\Psi_0} &= \coef{1}{2} \cos \frac{\pi}{8} \left( \ket{h}_a \ket{A_h}_a \ket{v}_b \ket{A_v}_b + \ket{v}_a \ket{A_v}_a \ket{h}_b \ket{A_h}_b \right) + \\
  &+ \coef{1}{2} \sin \frac{\pi}{8} \left( \ket{h}_a \ket{A_h}_a \ket{h}_b \ket{A_h}_b - \ket{v}_a \ket{A_v}_a \ket{v}_b \ket{A_v}_b \right).
\end{split}
\label{eq:pre}
\end{equation}
To proceed with step (ii), agent $E_A$ chooses between observables $A_0$ and $B_0$,
\begin{eqnarray}
  A_0 &=& \ket{h}_a \ket{A_h}_a \bra{h}_a \bra{A_h}_a - \ket{v}_a \ket{A_v}_a \bra{v}_a \bra{A_v}_a, \\
  B_0 &=& \ket{+}_A \ket{+}_A - \ket{-}_A \bra{-}_A,
\end{eqnarray}
where $\ket{\pm}_A = \left( \ket{h}_a \ket{A_h}_a \pm \ket{v}_a \ket{A_v}_a \right)/\sqrt{2}$. The first one, $A_0$, can be interpreted as a simple reading of agent's $I_A$ memory, whereas the second one, $B_0$, performs an external interference experiment, and therefore can be linked to agent's $E_A$ memory. Following the same spirit, agent $E_B$ chooses between $A_1$ and $B_1$,
\begin{eqnarray}
  A_1 &=& \ket{h}_b \ket{A_h}_b \bra{h}_b \bra{A_h}_b - \ket{v}_b \ket{A_v}_b \bra{v}_b \bra{A_v}_b, \\
  B_1 &=& \ket{+}_B \ket{+}_B - \ket{-}_B \bra{-}_B,
\end{eqnarray}
where $\ket{\pm}_B = \left( \ket{h}_b \ket{A_h}_b \pm \ket{v}_b \ket{A_v}_b \right)/\sqrt{2}$.

Finally, the third step is performed taking into account that
statements $1-4$ imply the existence of a joint probability
distribution $p(A_0,B_0,A_1,B_1)$ whose marginals satisfy the
Claude-Horne-Shimony-Holt (CHSH) inequality
\cite{Clauser:69,Zukowksi:14}
\begin{equation}
S=\left<A_1 B_1 \right> + \left<A_1 B_0 \right> + \left< A_0 B_1 \right> - \left< A_0 B_0 \right> \leq 2.
\label{eq:bell}
\end{equation}

In \cite{Bruckner:18} is theoretically shown that the initial state
given by Eq. (\ref{eq:bruckner2}) leads to $S=2\sqrt{2}$; in
\cite{Proietti:19} this result is confirmed by an experiment. The
conclusion is that these resuls are incompatible with statements
$1-4$, and therefore, assuming that statements $2$ (non-locality) and
$3$ (freedom of choice) are compatible with quantum mechanics
\cite{Bruckner:18,Proietti:19,Zukowksi:14}, quantum theory is
incompatible with the existence of observer-independent well
established facts.

\subsection{The role of the decoherence framework}

Unfortunately, this simple protocol is not consistent with the
decoherence framework. The previous analysis accounts neither for the
structure of laboratories summarized in Tab. \ref{tab:laboratorio},
nor for the measuring protocol given in Tab. \ref{tab:protocol}. The
decoherence framework postulates that a definite outcome does not
emerge until an external environment monitorizes the state composed by
the system, the apparatus and the observer. Therefore,
Eq. \eqref{eq:pre} does not represent the outcomes obtained by agents
$I_A$ and $I_B$, but just an entangled system composed by two photons
and two aparati. And consequently, the fact that it violates a CHSH
inequality does not entail the refutation of the fourth statement of
the theorem, since definite outcomes have not still appeared ---it
just shows that the state \eqref{eq:pre} includes quantum correlations
that cannot be described by means of a joint probability distribution,
but such correlations involve neither definite outcomes, nor
observers' memory records.

As a first conclusion, the previous paragraph is enough to show that
the thought experiment devised in \cite{Bruckner:18} cannot refute the
possibility of jointly assigning truth values to the propositions
about the outcomes of different observers, if the decoherence
framework is considered. Following the same line of reasoning that in
Sec. \ref{sec:discussion} we can also state that the conclusion in
\cite{Bruckner:18} is well supported if a measurement is understood as
a correlation between a system and its measuring apparatus. In such a
case, the correlations between $A$ and $B$ observables represent the
correlations between the outcomes obtained by the internal and the
external observables, and therefore the CHSH proves that they are
incompatible with a definite joint probability distribution.

The rest of the section is devoted to a variation of the setup devised
in \cite{Bruckner:18}. The idea is to follow the same spirit, but making it
suitable to challenge the decoherence framework. This modified experimental setup consists of two main steps:

(i) The internal agents measure their systems in the basis $\{ \ket{h}, \ket{v} \}$, and the external ones perform interference experiment in the basis $\{ \ket{+}, \ket{-} \}$.

(ii) Two super-external agents choose between the operators $A$ and
$B$, given by Eqs. (\ref{eq:a0})-(\ref{eq:b1}), to establish
complementary facts about the outcomes obtained in step (i).

This variation allows to apply a CHSH inequality to the outcomes
obtained by the internal and the external agents, and therefore to
test if we can jointly assing truth values to them. To properly apply
the decoherence framework to this experiment, it is mandatory to
include in the protocol all the environments which determine the
emergence of definite outcomes. This can be done in three different
stages:

{\em Stage 1.-} Agent $I_A$ measures the state of photon $a$ in a basis
given by $\left\{ \ket{h}_a, \ket{v}_a \right\}$, and $I_B$ measures
the state of photon $b$ in a basis given by $\left\{ \ket{h}_b,
\ket{v}_b \right\}$. Without explicitly taking into account the
external apparati and environments, which are not entangled with
laboratories $A$ and $B$ at this stage, the resulting state is
\begin{equation}
  \begin{split}
    \ket{\Psi_1} &= \coef{1}{2} \cos \frac{\pi}{8} \ket{h}_a \ket{A_h}_a \ket{\varepsilon_1(t)}_a \ket{v}_b \ket{A_v}_b \ket{\varepsilon_2(t)}_b + \\
    &+ \coef{1}{2} \cos \frac{\pi}{8} \ket{v}_a \ket{A_v}_a \ket{\varepsilon_2(t)}_a \ket{h}_b \ket{A_h}_b \ket{\varepsilon_1(t)}_b + \\
    &+  \coef{1}{2} \sin \frac{\pi}{8} \ket{h}_a \ket{A_h}_a \ket{\varepsilon_1(t)}_a \ket{h}_b \ket{A_h}_b \ket{\varepsilon_1(t)}_b - \\
    &- \coef{1}{2} \sin \frac{\pi}{8} \ket{v}_a \ket{A_v}_a \ket{\varepsilon_2(t)}_a \ket{v}_b \ket{A_v}_b \ket{\varepsilon_2(t)}_b.
  \end{split}
  \label{eq:case1}
\end{equation}
The decoherence framework establishes that agents $I_A$ and $I_B$ do
not observe definite outcomes until this stage is reached. It is worth
remembering that each of its environment is continuously monitorizing
each of its apparati, by means of Hamiltonians like
\eqref{eq:interaccion}.

At this stage, an experiment equivalent to the one discussed in
\cite{Bruckner:18,Proietti:19} could be done, by means of the
following $A_0$, $A_1$, $B_0$, and $B_1$ observables
\begin{subequations}
  \begin{align}
  \label{eq:a0}
  A_0 &= \ket{h}_a \ket{A_h}_a \bra{h}_a \bra{A_h}_a - \ket{v}_a \ket{A_v}_a \bra{v}_a \bra{A_v}_a, \\
  \label{eq:b0}
  B_0 &= \ket{h}_b \ket{A_h}_b \bra{h}_b \bra{A_h}_b - \ket{v}_b \ket{A_v}_b \bra{v}_b \bra{A_v}_b, \\
  \label{eq:a1}
  A_1(\tau) &= \ket{+(\tau)}_A \bra{+(\tau)}_A - \ket{-(\tau)}_A \bra{-(\tau)}_A, \\
  \label{eq:b1}
  B_1(\tau) &= \ket{+(\tau)}_B \bra{+(\tau)}_B - \ket{-(\tau)}_B \bra{-(\tau)}_B,
  \end{align}
\end{subequations}
where $\ket{+(\tau)}_A$ is given by Eq. (\ref{eq:mas});
$\ket{-(\tau)}_A$ is given by Eq. (\ref{eq:menos}), and equivalent
relations determine $\ket{+(\tau)}_B$ and $\ket{-(\tau)}_B$. Again,
$\tau$ is the time at which the interference measurements are
performed, according to points R1-R4 of Tab. \ref{tab:protocol}. In
such a case, $A_0$ and $B_0$ can be properly interpreted as agents'
$I_A$ and $I_B$ points of view, but $A_1$ and $B_1$ are still not
linked to agents' $E_A$ and $E_B$ perceptions ---their environments
must act to determine the corresponding definite outcomes. Hence, the
CHSH inequality applied to this state would allow us to get a
conclusion about the compatibility of the internal agents' memories and
the states of the laboratories in which they live, but they would tell
us nothing about the outcomes obtained by the external agents.

{\em Stage 2a.-} From Eq. (\ref{eq:case1}), agent $E_A$ measures the
state of laboratory $A$ in the basis $\left\{ \ket{+(\tau)}_A,
\ket{-(\tau)_A} \right\}$, considering requisites R1-R4 of
Tab. \ref{tab:protocol}. The resulting state is
\begin{equation}
  \begin{split}
    \ket{\Psi_2} &= \frac{1}{2} \left( \cos \frac{\pi}{8} - \sin \frac{\pi}{8} \right) \ket{+(\tau)}_A \ket{A'_+}_A \ket{\varepsilon'_1(\tau)}_A \ket{v}_b \ket{A_v}_b \ket{\varepsilon_2(\tau)}_b + \\
    &+ \frac{1}{2} \left( \cos \frac{\pi}{8} + \sin \frac{\pi}{8} \right) \ket{-(\tau)}_A \ket{A'_-}_A \ket{\varepsilon'_2(\tau)}_A \ket{v}_b \ket{A_v}_b \ket{\varepsilon_2(\tau)}_b + \\
    &+ \frac{1}{2} \left( \cos \frac{\pi}{8} + \sin \frac{\pi}{8} \right) \ket{+(\tau)}_A \ket{A'_+}_A \ket{\varepsilon'_1(\tau)}_A \ket{h}_b \ket{A_h}_b \ket{\varepsilon_1(\tau)}_b + \\
    &+ \frac{1}{2} \left( \sin \frac{\pi}{8} - \cos \frac{\pi}{8} \right) \ket{-(\tau)}_A \ket{A'_-}_A \ket{\varepsilon'_2(\tau)}_A \ket{h}_b \ket{A_h}_b \ket{\varepsilon_h(\tau)}_b.
  \end{split}
  \label{eq:case2}
\end{equation}
At this stage, observable $A_1$ represents agent's $E_A$ point of
view, but $B_1$ is still not linked to agent's $E_B$ memory.

{\em Stage 2b.-} From Eq. (\ref{eq:case1}) again, agent $E_B$ measures
the state of laboratory $B$ in the basis $\left\{ \ket{+(\tau)}_B,
\ket{-(\tau)_B} \right\}$, considering requisites R1-R4 of
Tab. \ref{tab:protocol}. The resulting
state is
\begin{equation}
\begin{split}
  \ket{\Psi_3} &= \frac{1}{2} \left( \cos \frac{\pi}{8} - \sin \frac{\pi}{8} \right) \ket{v}_a \ket{A_v}_a \ket{\varepsilon_2(\tau)}_a \ket{+(\tau)}_B \ket{A'_+}_B \ket{\varepsilon'_1 (\tau)}_B + \\
  &+ \frac{1}{2} \left( \cos \frac{\pi}{8} + \sin \frac{\pi}{8} \right) \ket{v}_a \ket{A_v}_a \ket{\varepsilon_2(\tau)}_a \ket{-(\tau)}_B \ket{A'_-}_B \ket{\varepsilon'_2 (\tau)}_B + \\
  &+ \frac{1}{2} \left( \cos \frac{\pi}{8} + \sin \frac{\pi}{8} \right) \ket{h}_a \ket{A_h}_a \ket{\varepsilon_1(\tau)}_a \ket{+(\tau)}_B \ket{A'_+}_B \ket{\varepsilon'_1 (\tau)}_B + \\
  &+ \frac{1}{2} \left( \sin \frac{\pi}{8} - \cos \frac{\pi}{8} \right) \ket{h}_a \ket{A_h}_a \ket{\varepsilon_1(\tau)}_a \ket{-(\tau)}_B \ket{A'_-}_B \ket{\varepsilon'_2 (\tau)}_B.
\end{split}
\label{eq:case3}
\end{equation}
As this stage has been obtained from Eq. \eqref{eq:case1}, it is not
subsequent to Eq. \eqref{eq:case2}. Therefore, observable $B_1$
represents agent's $E_B$ point of view, but $A_1$ is still not linked
to agent's $E_A$ memory.

{\em Stage 3.-} Agent $E_A$ measures the state of laboratory $A$ in the
basis $\left\{ \ket{+(\tau)}_A, \ket{-(\tau)_A} \right\}$, considering
requisites R1-R4 of Tab. \ref{tab:protocol}, and agent $E_B$ measures
the state of laboratory $B$ in the basis $\left\{ \ket{+(\tau)}_B,
\ket{-(\tau)_B} \right\}$, following the same procedure. This case is
subsequent to either stage 2a or stage 2b. The resulting state is
\begin{equation}
  \begin{split}
    \ket{\Psi_4} &= \coef{1}{2} \cos \frac{\pi}{8} \ket{+(\tau)}_A \ket{A'_+}_A \ket{\varepsilon'_1(\tau)}_A \ket{+(\tau)}_B \ket{A'_+}_B \ket{\varepsilon'_1(\tau)}_B - \\
    &- \coef{1}{2} \cos \frac{\pi}{8} \ket{-(\tau)}_A \ket{A'_-}_A \ket{\varepsilon'_2(\tau)}_A \ket{-(\tau)}_B \ket{A'_-}_B \ket{\varepsilon'_2(\tau)}_B + \\
    &+ \coef{1}{2} \sin \frac{\pi}{8} \ket{+(\tau)}_A \ket{A'_+}_A \ket{\varepsilon'_1(\tau)}_A \ket{-(\tau)}_B \ket{A'_-}_B \ket{\varepsilon'_2(\tau)}_B + \\
    &+ \coef{1}{2} \sin \frac{\pi}{8} \ket{-(\tau)}_A \ket{A'_-}_A \ket{\varepsilon'_2(\tau)}_A \ket{+(\tau)}_B \ket{A'_+}_B \ket{\varepsilon'_1(\tau)}_B. \\
  \end{split}
  \label{eq:case4}
  \end{equation}

At this stage, the four agents have observed definite outcomes, and
therefore their four memories can be read to interpret these
outcomes. Hence, if one wants to test statements 1-4 of the
theorem formulated in \cite{Bruckner:18}, one must start from this
state. So, let us image that we are running a quantum algorithm performing
this experiment and we want to test if we can jointly assign truth
values to the outcomes obtained by the four agents. One of us can
choose between $A_0$ and $B_0$, defined in Eqs. \eqref{eq:a0} and
\eqref{eq:b0}, to decide between reading agent's $I_A$ or agent's
$E_A$ memories, and another one can choose between $A_1$ and $B_1$,
defined in Eqs. \eqref{eq:a1} and \eqref{eq:b1}, to decide between reading agent's $I_B$ or agent $E_B$ memories. Then, we can run a large
number of realizations of the same experiment to test if the CHSH
inequality given in Eq. \eqref{eq:bell} holds. If it gives rise to
$S>2$ we can conclude that statement (4) of the theorem is violated;
if not, we can conclude that it is possible to jointly assign truth
values to the observations done by the four agents.

A straightforward calculations provides the following result,
\begin{eqnarray}
  \expected{\Psi_4}{A_0 B_0}{\Psi_4} &= 0, \\
  \expected{\Psi_4}{A_1 B_0}{\Psi_4} &= 0, \\
  \expected{\Psi_4}{A_0 B_1}{\Psi_4} &= 0, \\
  \expected{\Psi_4}{A_1 B_1}{\Psi_4} &= 1/\sqrt{2}.
\end{eqnarray}
Therefore, the CHSH inequality, Eq. \eqref{eq:bell}, applied to $\ket{\Psi_4}$ leads to
$S=1/\sqrt{2}<2$.

Two main conclusions can be gathered from this section. First, the
experiment devised in \cite{Bruckner:18}, and its experimental
realization \cite{Proietti:19}, are incompatible with the decoherence
framework, because, according to it, they do not deal with proper
outcomes; thus, they cannot be used to refute the possibility of
jointly assigning truth values to the outcomes obtained by different
observers. Second, a variation to the experiment in \cite{Bruckner:18}
designed to challenge the decoherence framework following the same
spirit is compatible with the four statements discussed above
---quantum theory is valid at any scale; the choice of the measurement
settings of one observer has no influence on the outcomes of other
distant observers; the choice of the measurement settings is
independent form the rest of the experiment, and one can jointly
assign truth values to the propositions about the outcomes of
different observers. In other words, these statements do not imply a
contradiction in this experiment, if the role of all the parts of each
laboratory, given in Tab. \ref{tab:laboratorio}, and the physical
mechanisms giving rise to each outcome, are considered. This
conclusion is fully compatible with the main idea behind the
decoherence framework. As it is clearly stated in the title of
Ref. \cite{Zurek:03}, the main aim of this formalism is to explain how
classical results, like definite outcomes, can be obtained from
quantum mechanics, without relying on a non-unitary wave-function
collapse. This section shows that the memory records of all the agents
are indeed classical as a consequence of the continuous monitorization
by their environments, and therefore satisfy the corresponding CHSH
inequality. Notwithstanding, the state after all these definite
outcomes have emerged, Eq. \eqref{eq:case4}, is quantum and has true
quantum correlations. If one applies the CHSH inequality to the states
of the two internal and the two external laboratories, the resulting
equations are formally the same that those in \cite{Bruckner:18}
---confirming that the thought experiment discussed in this section
follows the same spirit that the one in \cite{Bruckner:18}---, and
therefore one recovers the original result, $S=2 \sqrt{2}$. This means
that we cannot assign joint truth values to the state of these
laboratories, but we can make this assignement to the state of the
agents memories. In \cite{Bruckner:18,Proietti:19} there is no
distinction between the state of the laboratory in which an agent
lives, and the state of its memory; the decoherence framework is based
precisely on this distinction.

Before ending this section, it is worth remarking that our
result does not prove that the use of statements $1-4$ is free from
contradictions in any circumstances. We have just shown that the
particular setup used to prove the no-go theorem in \cite{Bruckner:18}
does not lead to contradictions if the decoherence framework is
properly taken into account. But, again, the main statement of the
theorem can be still considered as a conjecture.

  \section{Conclusions}
  \label{sec:conclusiones}

The main conclusion of this work is that neither the original Wigner's
friend experiment, nor the extended version proposed in
\cite{Renner:18}, nor the one in \cite{Bruckner:18} (and its
corresponding experimental realization, \cite{Proietti:19}) entail
contradictions if the decoherence framework is properly taken into
account.

This framework consists in considering that a quantum measurement and
the corresponding (apparent) wave-function collapse are the
consequence of the interaction between the measuring apparatus and an
uncontrolled environment, which must be considered as an inseparable
part of the measuring device. In this work, we have relied on a simple
model to show that a chaotic interaction is necessary to induce such
an apparent collapse, but, at the same time, a quite small number of
environmental qbits suffices for that purpose. This implies that any
experiment on any quantum system can be modeled by means of a unitary
evolution, and therefore all the time evolution, including the
outcomes obtained by any observers, is univocally determined by the
initial state, the interaction between the system and the measuring
apparati, and the interaction between such apparati and the
corresponding environments. Seeing the reality as if a random
wave-function collapse had happend is due to the lack of information
suffered by the observers ---only the system as a whole evolves
unitarilly, not a part of it. This is a somehow paradoxical solution
to the quantum measurement problem: ignoring an important piece of
information about the state in which the observer lives is mandatory
to observe a definite outcome; taking it into account would lead to no
observations at all. But, besides the ontological problems arising for
such an explanation, the resulting framework is enough for the purpose
of this work.

The first consequence of this framework is that the memory records of
Wigner's friends ---the internal agents in a Wigner's friend
experiment--- change as a consequence of the external interference
experiment performed by Wigner, these changes are univocally
determined by the Hamiltonian encoding all the time evolution, and
therefore can be exactly predicted. Hence, if an agent has
observed a definite outcome, then external interference experiments
performed on the laboratory in which it lives change its memory
records; if such changes do not occur is because the agent has not
observed a definite outcome.

The second consequence of the decoherence framework is that the
contradictions discussed in \cite{Renner:18} and \cite{Bruckner:18}
are ruled out. If the agents involved in the thought experiment
devised in \cite{Renner:18} use the decoherence framework as the
common theory to predict the other agents' outcomes, their conclusions
are not contradictory at all. The analysis of the experiment proposed
in \cite{Bruckner:18} is a bit more complicated. Its original design
is not compatible with the decoherence framework. Hence, it cannot be
used to refute the possibility to jointly assigning truth values to
the agents' outcomes, that is, it cannot be used to prove the no-go
theorem formulated in \cite{Bruckner:18}. Therefore, a variation of
that experiment, following the same spirit, is proposed to show that,
if the CHSH inequality is applied to the state at which the whole
system is at the end of the protocol, that is, when the records in the
memories of the four agents are fixed, the resulting value is
compatible with the existence of observer-independent facts.

However, this is not enough to dismiss the main statements of the
no-go theorems formulated in such references. The conclusion of this
work is that the examples used to prove these theorems are not valid
within the decoherence framework, but we have not proved that this
framework is totally free from similar inconsistencies. Hence, these
statements can be still considered as conjectures. Further work is
required to go beyond this point.

It is also worth to remark that the decoherence formalism also narrows
down the conditions under which the external interference
measurements, trademark of Wigner's friend experiments, are expected
to work. This means that, if the decoherence framework results to be
true, human beings acting as observers are almost free from suffering
the strange effects of such experiments. Notwithstanding, the
promising state-of-the-art in quantum technologies may provide us, in
the future, quantum machines able to perform these experiments.

Finally, the conclusion of this work must not be understood as a
strong support of the decoherence framework. It just establishes that
such a framework does not suffer from the inconsistencies typically
ensuing Wigner's friend experiments. However, there is plenty of
space for theories in which the wavefunction collapse is real
\cite{Bassi:13}. These theories predict a totally different scenario,
since after each measurement the wave function of the whole system
collapses, and therefore becomes different from the predictions of the
decoherence framework. Hence, experiments like the ones discussed in
this work might be a way to test which of this proposals is correct
---if any.

\begin{acknowledgements}
This work has been supported by the Spanish Grants
Nos. FIS2015-63770-P (MINECO/ FEDER) and PGC2018-094180-B-I00
(MCIU/AEI/FEDER, EU). The author acknowledges A. L. Corps for his
critical reading of the manuscript.
\end{acknowledgements}


\begin{thebibliography}{99}
\bibitem{nota0} One of the trademarks of the decoherence
      interpretation of quantum mechanics is that the pointer states
      of any apparati, that is, the states which appear as objective
      outcomes, are those which survive to the continuous monitorization
      by a complex environment. In other words, the environment, and its
      interaction with the measuring apparatus, is the origin of the
      classical perception of the reality. For a more detailed
      discussion, we refer the reader to \cite{Zurek:03,Zurek:07}.
\bibitem{nuevanota} Of course, the agent cannot restore the complete state from a single measurement. If its outcome is, say, $h$, it can just conclude that the global state must be $\ket{\Psi_2} = \alpha \ket{h} \ket{A_h} \ket{\epsilon_1(t)} + \beta \ket{v} \ket{A_v} \ket{\epsilon_2(t)}$, with unknown coefficents $\alpha$ and $\beta$ such that $\left| \alpha \right|^2 + \left| \beta \right|^2 = 1$, and $\left| \alpha \right|>0$. It is very important to note that this conclusion, which is one of the trademarks of the decoherence formalism, is totally different from the standard interpretation of quantum measurements, following which the observer concludes that the state of the system is $\ket{\Psi_2} = \ket{h} \ket{A_h}$ as the aftermath of observing the outcome $h$.
\bibitem{nota5} From this point of view, we can consider that the
    measurement is done by a small quantum machine, and that an
    amplification process is done by the environment afterwards. In
    this way, the pre-measurement can be considered purely quantum,
    whereas the amplification, required for a large agent to see the
    outcome, is the responsible for the transition to a {\em
      classical} outcome. Nevertheless, the result is the same if the
    external environment-apparatus interaction is still present,
    provided that its characteristic time is larger than the time
    required to complete the pre-measurement.
\bibitem{nota_colapso} An interesting question is what happens if the
  non-diagonal elements, $C_{hv}$ and $C_{vh}$, are clearly different
  from zero. In such a case, the interaction between the apparatus and
  the environment does not determine the pointer states of the
  apparatus, and therefore the agent cannot see a definite outcome. On
  the contrary, the instantaneous eigenbasis of the reduced density
  matrix resulting from tracing out the environmental degrees of
  freedom is continuously changing in time, and therefore the agent
  memory does not record a definite result for the measurement. As the
  fluctuations of these diagonal elements are expected to be tiny if
  the environment is large enough, the outcomes can be considered
  stable. A more detailed discussion can be found in \cite{Zurek:03}.
\end{thebibliography}
\end{document}